\documentclass[twoside,twocolumn,9pt]{article}
\usepackage{extsizes}
\usepackage[sort&compress,comma,super]{natbib} 
\usepackage[version=3]{mhchem}
\usepackage[left=1.5cm, right=1.5cm, top=1.785cm, bottom=4.0cm]{geometry}
\usepackage{balance}
\usepackage{times,mathptmx}
\usepackage{sectsty,siunitx}
\usepackage{subfig}
\usepackage{graphicx} 
\usepackage{booktabs}
\usepackage{lastpage}
\usepackage[format=plain,justification=justified,singlelinecheck=false,font={stretch=1.125,small,sf},labelfont=bf,labelsep=space]{caption}
\usepackage{float}
\usepackage{fancyhdr}
\usepackage{extramarks}
\usepackage{afterpage,lastpage}
\usepackage{fnpos}
\usepackage[english]{babel}
\addto{\captionsenglish}{%
  
}
\usepackage{array}
\usepackage{droidsans}
\usepackage{charter}
\usepackage{color, colortbl}
\usepackage[T1]{fontenc}
\usepackage[usenames,dvipsnames]{xcolor}
\usepackage{setspace}
\usepackage[compact]{titlesec}
\usepackage{hyperref}
\usepackage{soul}
\usepackage{physics}

\newcommand\TinkerHP{Tinker-\textit{HP}}

\newcommand{\PyCUDA}{{\textsf{PyCUDA}}}
\newcommand{\PyTorch}{{\textsf{PyTorch}}}
\newcommand{\TensorFlow}{\textsf{TensorFlow}}
\newcommand{\TorchANI}{{\textsf{TorchANI}}}
\newcommand{\cffi}{\textsf{cffi}}
\newcommand{\Fortran}{{\textsf{Fortran}}}
\newcommand{\C}{{\textsf{C}}}
\newcommand{\Python}{{\textsf{Python}}}

\newcommand{\GPU}{\emph{GPUs}}
\newcommand{\gpu}{\emph{GPU}}

\newcommand{\cpu}{\emph{CPU}\ }

\newcommand{\ttiny}[1]{\text{\tiny{#1}}}

\usepackage{epstopdf}

\definecolor{Gray}{gray}{0.9}
\definecolor{cream}{RGB}{222,217,201}
\definecolor{caputmortuum}{rgb}{0.35, 0.15, 0.13}

\fancypagestyle{ici}{
\fancyhf{}

}

\fancypagestyle{ici2}{
\fancyhf{}

}



\makeFNbottom
\makeatletter
\renewcommand\LARGE{\@setfontsize\LARGE{15pt}{17}}
\renewcommand\Large{\@setfontsize\Large{12pt}{14}}
\renewcommand\large{\@setfontsize\large{10pt}{12}}
\renewcommand\footnotesize{\@setfontsize\footnotesize{7pt}{10}}
\makeatother

\makeatletter 
\renewcommand\@biblabel[1]{#1}            
\renewcommand\@makefntext[1]%
{\noindent\makebox[0pt][r]{\@thefnmark\,}#1}
\makeatother 

\sectionfont{\sffamily\Large}
\subsectionfont{\normalsize}
\subsubsectionfont{\bf}
\titlespacing*{\section}{0pt}{4pt}{4pt}
\titlespacing*{\subsection}{0pt}{15pt}{1pt}

\makeatletter 
\newlength{\figrulesep} 
\makeatother

\begin{document}

\twocolumn[
  \begin{@twocolumnfalse} 

\vspace{3cm}
\sffamily
\begin{tabular}{m{4.5cm} p{13.5cm}}

\includegraphics{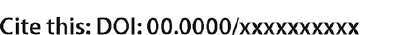} & \noindent\LARGE{\textbf{Scalable Hybrid Deep Neural Networks/Polarizable Potentials Biomolecular Simulations including Long-range Effects$^\dag$}} \\

{} & \noindent\large{Théo Jaffrelot Inizan,\textit{$^{a}$} Thomas Plé,\textit{$^{a}$} Olivier Adjoua,\textit{$^{a}$} Pengyu Ren,\textit{$^{b}$} Hatice Gökcan,\textit{$^{c}$} Olexandr Isayev,\textit{$^{c}$} Louis Lagardère,\textit{$^{a,d}$} and Jean-Philip Piquemal\textit{$^{a,b}$}*} \\

\includegraphics{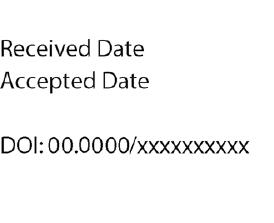} & \noindent\normalsize{Deep-HP is a scalable extension of the \TinkerHP\ multi-GPUs molecular dynamics (MD) package enabling the use of Pytorch/TensorFlow Deep Neural Networks (DNNs) models. Deep-HP increases DNNs MD capabilities by orders of magnitude offering access to ns simulations for 100k-atom biosystems while offering the possibility of coupling DNNs to any classical (FFs) and many-body polarizable (PFFs) force fields. It allows therefore to introduce the ANI-2X/AMOEBA hybrid polarizable potential designed for ligand binding studies where solvent-solvent and solvent-solute interactions are computed with the AMOEBA PFF while solute-solute ones are computed by the ANI-2x DNN. ANI-2X/AMOEBA explicitly includes AMOEBA’s physical long-range interactions via an efficient Particle Mesh Ewald implementation while preserving ANI-2X’s solute short-range quantum mechanical accuracy. The DNNs/PFFs partition can be user-defined allowing for hybrid simulations to include biosimulation key ingredients such as polarizable solvents, polarizable counter ions, etc... ANI-2X/AMOEBA is accelerated using a multiple-timestep strategy focusing on the models contributions to low-frequency modes of nuclear forces. It primarily evaluates AMOEBA forces while including ANI-2x ones only via correction-steps resulting in an order of magnitude acceleration over standard Velocity Verlet integration. Simulating more than 10 $\mu$s, we compute charged/uncharged ligands solvation free energies in 4 solvents, and absolute binding free energies of host–guest complexes from SAMPL challenges. ANI-2X/AMOEBA average errors are within chemical accuracy opening the path towards large-scale hybrid DNNs simulations, at force-field cost, in biophysics and drug discovery.

}
\end{tabular}

 \end{@twocolumnfalse} 

  ]

\renewcommand*\rmdefault{bch}\normalfont\upshape
\rmfamily
\section*{}
\vspace{-1cm}


\footnotetext{\textit{$^{a}$~Sorbonne Université, Laboratoire de Chimie Théorique, UMR 7616 CNRS, 75005, Paris, France}}
\footnotetext{\textit{$^{b}$~University of Texas at Austin, Department of Biomedical Engineering, Austin, Texas, USA}}
\footnotetext{\textit{$^{c}$~Carnegie Mellon University, Department of Chemistry, Pittsburgh, Pennsylvania, USA.}}
\footnotetext{\textit{$^{d}$~Sorbonne Université, Institut Parisien de Chimie Physique et Théorique, FR 2622 CNRS, Paris, France}}

\footnotetext{\textit{$^{\*}$~Contact:  jean-philip.piquemal@sorbonne-universite.fr }}

\footnotetext{\dag~Electronic Supplementary Information (ESI) available: See DOI: 00.0000/00000000.}





\pagestyle{ici}

\newpage

\section{Introduction}
Understanding the dynamics of biological systems is of prime importance in structural biology and drug discovery. Over the last 50 years, coupled to force fields (FFs), molecular dynamics (MD) simulations have proven to be an essential theoretical tool to predict the long-timescale behaviour of proteins in complex environments. In recent years, deep learning technologies have also progressed and showed some potential to accelerate drugs discovery. For example, in the last months, DeepMind developed the Alphafold² \citep{alphafold2} model that is able to predict over 200 million protein structures. Proteins' properties could, however, drastically change during a molecular dynamics simulation. For instance, the protein-water interface can drive fluctuations of catalytic cavities and thus change drug inhibition. MD is therefore the prominent approach to go beyond simple structure in order to predict the complete protein conformational space.\citep{proteaseadaptive,interfacial_dina,D1SC05892D} Due to the biological system sizes and biological simulation timescales, pure quantum chemistry models cannot be used for simulations and are replaced by empirical FFs, that are presently commonly used to model chemical interactions. \\
FFs model the total energy as a sum over intra and intermolecular energy terms. The treatment of the latter leads to two classes of FFs: classical and polarizable. In classical FFs, the intermolecular interactions are modeled by Lennard-Jones and Coulomb potential which make them computationally efficient enabling modern softwares to tackle long timescale simulation of complex systems.\cite{NAMD,ANTON,GROMACS,GENESIS} While offering a reasonable precision thanks to careful parametrization, \cite{PONDER200327,Monticelli2013}, classical FFs lack an accurate description of polarization and to a larger extent of many-body physical effects.\cite{NohadGresh2007,10.3389/fmolb.2019.00143} These quantities can play a crucial role in solvation \cite{proteaseadaptive,interfacial_dina} and for the stability of secondary, and quaternary structures of proteins.\cite{10.3389/fmolb.2019.00143} 
The development of polarizable FFs (PFFs) has opened new routes able to explicitly include many-body effects.\citep{chapterpol,doi:10.1146/annurev-biophys-070317-033349} 
Their computational cost has long hindered their use but with the rise of High Performance Computing (HPC)\cite{tinkerhp,Drude} and the increasing performance of computational devices such as \GPU, million-atoms PFFs simulations are now possible.\cite{tinkerhpgpu} \\
At this stage, Machine Learning (ML) schemes have also the potential to offer a new paradigm for boosting MD simulations and to take their part in the development of FFs. 
ML potentials (MLP) also avoid solving the Schrödinger equation at each time-step of the simulation by providing a mathematical direct relation between the atomic positions and the potential energy. 
In recent years, MLPs have been an active field of research which led to the emergence of different framework such as high-dimensional deep neural network potentials (HDNNPs), Gaussian approximation potentials (GAP),\citep{GAP_gabor} moment tensor potentials, spectral neighbor analysis potentials (SNAP),\citep{SNAP} atomic cluster expansion, graph networks, kernel ridge regression methods,\citep{Kernel_ridge_regression_vovk} gradient-domain machine learning (GDML)\citep{GDML, sGDML_1, sGDML_2, sGDML_3} and support vector machines (SVM).\citep{SVM} 
MLP nonlinear functional forms are very general and highly flexible, allowing for a very accurate representation of electronic structure computations reference data. 
The input of a MLP is usually a hand-crafted real valued functions of the coordinates that preserve some symmetries and uniquely defined atomic environments. In practice, the choice of this descriptor is central to design an accurate MLP. A variety of physics-based descriptor have been developed such as the smooth overlap of atomic positions (SOAP),\citep{SOAP} the spectrum of approximated Hamiltonian matrices representations (SPAHM),\citep{SPAHM} the Coulomb matrix (CM) and the atom-centered symmetry functions (ACSFs).\citep{HDNNP_behlerpar, wACSF} The latter, introduced by Behler and Parinello in 2007, is still the most popular descriptor used for HDNNP and have been employed in numerous studies.\citep{4gen_HDNNP,HDNNP_behlerpar} It describes the atomic environment of a given central atom inside a cutoff radius $R_c$ by the use of radial and angular functions. Some modifications of the initial symmetry functions have been done since, aiming to reduce the number of symmetry functions that exhibit quadratic growth with the number of elements or improve the probing of the atomic environment\citep{ANI1X}. However, even if such descriptor have considerably improved the transferability and the scalability of HDNNPs, they are often use to only study small chemical systems that remain far away from the needs of biological modeling. They have nevertheless already be shown to be useful to create buffer region neural network in QM/MM (Quantum Mechanics/Molecular Mechanics) simulations to minimize overpolarization artifacts of the QM region due to classical MM.\cite{buffernn}  Another issue has been the lack of efficient MLP multi-GPU infrastructure software inside an already existing molecular dynamics package. 
In the last couple of years things started to change and our work is part of this large movement and also aims to address recent development of the ML-field.\citep{pushing_limit_car} 
While our work aims to utilize new developments in ML-field, we also aim to address some of the shortcomings of MLPs. Indeed, the intrinsic architecture of MLP usually constrains them to short-range interactions. Recently, Tsz Wai Ko \textit{et al.} proposed a fourth-generation of HDNNP which is able to capture long-range charge transfer and multiple charge states.\citep{4gen_natcom_behler} While it demonstrates the power of ML, their computational cost is much higher compared to physics-based PFFs long-range models and is not yet able to correctly describe solute in water.\\ To address these challenges, we present Deep-HP (HP stands for High-Performance), a multi-GPU MLP platform which is part of the \TinkerHP\ package and enables the coupling and development of MLP with state-of-the-art many-body polarizable effects. \TinkerHP\ uses massive parallelization by means of 3D decomposition which is a particularly well suited strategy for MLPs that are often developed by decomposing the total energy as a sum of atomic energy contributions.\citep{tinkerhp,tinkerhpgpu} 
The platform theoretical scalability with MLPs is linear and allows to scale up to hundreds/thousands of \GPU on large systems. 
As the present code shares the \TinkerHP\ capabilities, it allows 
for invoking fast physics-based many-body energy contributions. We extensively test Deep-HP scalability and implementation on the ANI model, one of the most accurate MLPs to date for small organic molecules. Finally, in the spirit of polarizable  QM/MM embedding simulations,\cite{AMOEBAQMMD,polembedd,C9SC01745C} we introduce an hybrid DNN/MM strategy that uses the ANI DNN to model solute-solute interactions and the AMOEBA PFF to evaluate solvent-solute and solvent-solvent interactions. This enables ANI to benefit from AMOEBA's strenghts that include an accurate condensed phase flexible water and protein models, the capability to include counter-ions and long-range/many-body effects. It should increase ANI transferability to a broader range of systems including charged ones. The performance of the model is evaluated by calculating the solvation free energies of various molecules in four organic solvents as well as the binding free energies of 14 challenging host-guest complexes taken from SAMPL blind challenges.

\section{Method}
\subsection{Potential Energy Models}

\subsubsection{The AMOEBA Polarizable Force Field}
The total potential energy of the AMOEBA \cite{ren2003polarizable,status_amoeba} polarizable model is expressed as the sum of bonded and non-bonded energy terms:
\begin{equation}
    \begin{split}
   & E_{total}=E_{bonded}+E_{non-bonded} \\
   & E_{bonded}=E_{bond}+ E_{angle}+E_{b\theta} \\
   & E_{non-bonded}=E_{vdW}+E_{ele}^{perm}+E_{ele}^{pol}
    \end{split}
\end{equation}
The bonded terms embody MM3-like ~\cite{allinger1989molecular} anharmonic bond-stretching and  angle-bending terms. Regarding the specific case of the polarizable AMOEBA water model, the intramolecular geometry and vibrations are described with an Urey-Bradley approach.\cite{ren2003polarizable}

The non-bonded terms include the van der Waals (vdW) interactions and the electrostatic contributions from both permanent and induced dipoles (polarization). More precisely, the polarization contribution is computed using an Applequist/Thole model \cite{thole1981molecular} whereas Halgren's buffered 14-7 pair potential is used to model vdW interactions.\cite{halgren1992representation} 
Computing the polarization energy requires the resolution of a linear system to get the induced dipoles, which is made through the use of iterative solvers such as preconditionned conjugated gradient that is the one used in this paper (with a \num{1e-5} tolerance).\cite{tinkerhpgpu}

To model the electrostatic interactions, AMOEBA relies on point atomic multipoles truncated at the quadrupole level. More details about the functional form and parametrization of AMOEBA can be found in reference.\cite{poltype2}
Electrostatics and many-body polarization long-range interactions are fully included through the use of the Smooth Particle Mesh Ewald approach \cite{SPME,lagardere2015scalable} that allows for efficient n(log(n)) periodic boundary conditions simulations. Beside water \cite{ren2003polarizable}, AMOEBA is a general force field available for many solvent \cite{ponder2010current}, ions,\cite{ionmonoAMOEBA, iondivAMOEBA} proteins \cite{Shi} and nucleic acids \cite{amoeba_nucl} biomolecular simulations.

\subsubsection{Neural Network Potentials}
Feed-forward neural network (FFNN) is a machine learning model that uses as building blocks connected layers of nodes (i.e neurons) 
each associated with their weights and bias. The output of each neuron is computed through a function of the output of the previous layer. Each weight is the strength associated to a specific node connection and they are updated during the training process. The depth (i.e number of layers) of the FFNN is related to its flexilibity and the complexity of the training dataset. Through careful optimization of hyperparameters, weights, biases and architecture, the FFNN can learn high dimensional non-linear functions such as potential energy surfaces (PESs). For HDNNP, the FFNN maps molecular structures to potential energy. The original HDNNP, introduced by Behler and Parrinello, expresses the total energy of a system $E_T$ as a sum of atomic contributions $E_i$.
\begin{equation}
\begin{aligned}
    E_T = \sum_{i}^{N_{atoms}} E_i\qty(G_i)\label{eq:energy_decomposition}
\end{aligned}
\end{equation}
where $G_i$ is the atomic environment vector (AEV) of atom $i$.
Based on the assumption of locality, each atom $i$ is associated with an AEV which probes specific radial and angular chemical regions. Each $G_{i}$ is then used as input into a single HDNNP. 
The construction of AEVs for each atom in the system enable the use of models for large systems even though they are trained on small molecules.   
Moreover, this summation has the advantage that it scales linearly with respect to the number of atoms. This atomic decomposition scheme has notably accelerated the development of HDNNP with increasingly complex architectures and AEV schemes.

\subsubsection{ANI models}
Smith et al. developed ANI, a model that uses a modified version of the Behler-Parinello symmetry functions (BPSFs).\citep{less_is_more,ANI1X} Symmetry functions are building blocks of the so-called atomic environment vector (AEV), $G_{i} = \{G_{1}^{X}, ..., G_{M}^{X}\}$, which aims to probe angular and radial local environment of a central atom $i$ with atomic number $X$. The locality approximation is achieved by using a differentiable cutoff function:
\begin{equation}
    f_{c}(R_{ij}) = \left\lbrace
        \begin{array}{ll}
            0 \hspace{3.1cm} & R_{ij} > R_c\\
            \dfrac{1}{2}\cos{\{\frac{\pi R_{ij}}{R_c}\}}+0.5 & R_{ij} \leq R_c
        \end{array} \right.
\end{equation}
where $R_{ij}$ is the distance between the central atom $i$ and a neighbor $j$, and $R_c$ a cutoff radius, here fixed to \SI{5.2}{\angstrom}. To probe the neighboring environment of the central atom inside the cutoff sphere, the AEV is divided into two types of symmetry functions: radial and angular. \\
The commonly used radial function is a sum of products of Gaussian and cutoff functions as introduced by Behler-Parinello:
\begin{equation}
    G^{rad}_{i,m} = \sum_{j \neq i}^{N_{atoms} \in R_{c}} e^{-\eta(R_{ij}-R_{s})^{2}}f_{c}(R_{ij})
\end{equation}
The index $m$ is associated to a set of parameters $\{ \eta, R_s \}$, where $R_s$ is the distance from the central atom for which the center of the Gaussian is shifted and $\eta$ is the spatial extension of the Gaussian. \\
The radial symmetry functions are not sufficient to distinguish between chemical environment, e.g if the neighboring atoms are all at the same distance from atom $i$. This is solved by using angular symmetry functions,
\begin{equation}
    G_{ANI-ang}^{i,m} = 2^{1-\xi} \sum_{j,k \neq i}^{N_{atoms}} (1+\cos{(\theta_{ijk}-\theta_{s})})^{\xi}e^{-\eta(\frac{R_{ij}+R_{ik}}{2}-R_{s})^{2}}f_{c}(R_{ij})f_{c}(R_{ik})
\end{equation}
where $\theta_{ijk}$ is the angle between the central atom $i$ and neighbors $j$ and $k$, $\theta_s$ used to center the maxima of the cosine and $\xi$ changes the width of the peak. To differentiate between atom species, ANI supplied a radial part for each atomic number and an angular part for each corresponding pair inside the cutoff sphere $R_c$. Thus, for $N$ atom species, the AEV has $N$ radial and $\frac{N(N+1)}{2}$ angular sub-AEVs. \\
The first ANI potential, ANI-1x \citep{ANI1_data, ANI1_fulldata}, 
has been developed for simulating organic molecules containing H, C, N, and O chemical elements.
The recent extension to ANI, ANI-2x\citep{ANI2X},has been trained to three additional chemical elements (S, F, and Cl).
This model extends the capabilities of ANI towards more diverse chemical structures such as proteins that often contain Sulfur and Chlorine atoms.\citep{ANI2X} \\
As ANI remains mainly designed to study the dynamics of small- to medium-size organic molecules, it had not been initially coupled to a massively parallel infrastructure. In contrast, another popular MLP, introduced by Weinan et al. \citep{DeePMD_water, DeePMD-kit}, DeePMD has been pushed towards large scale simulations of millions of atoms but has been trained on some specific systems, limiting its transferability. 

\subsubsection{DeePMD Models} 
The specificity of DeePMD compared to other MLPs is that it does not use hand-crafted symmetry functions to get the atomic environment\citep{DeePMD_water, DeePMD-kit}. \\
For an atom $i$, its $j$ neighbors within a cutoff radius are first sorted according to their chemical species and their inverse distances to the central atom. \\
The central atom is then associated to its local frame ($e_{x}$, $e_{y}$, $e_{z}$) and the local coordinates of its neighbors is denoted ($x_{ij}$, $y_{ij}$, $z_{ij}$). The atom $i$ local environment $\{D_{ij}\}$ is then defined as:

\begin{equation}
\begin{aligned}
    \{D_{ij}\} = \left\lbrace \frac{1}{R_{ij}}, \frac{x_{ij}}{R_{ij}}, \frac{y_{ij}}{R_{ij}}, \frac{z_{ij}}{R_{ij}}\right\rbrace
\end{aligned}
\end{equation}

$\{D_{ij}\}$ is then used as input for a FFNN to predict the atomic energy $E_{i}$. \\
DeePMD has been recently pushed in order to simulate tens of millions atoms for water and copper using a highly optimized \gpu\ code on the Summit supercomputer \citep{pushing_limit_car} but it would hugely benefit from all the available feature of \TinkerHP\ in order to run large scale biological simulations.

\subsubsection{Hybrid Model: Neural Network Solutes in AMOEBA Polarizable Solvent/Protein}

Hybrid DNN/MM simulations using classical FFs have been introduced by Lahey and Rowley .\cite{Rowley_binding}
One technical issue with hybrid DNN/FF approaches is that in local MLP models such as ANI and DeePMD, each atom only interacts with its closest neighbors within a relatively small cutoff radius. Therefore a correct description of long-range interactions is crucial for the simulation of condensed-phase systems, making them particularly challenging for MLP models.\cite{norberg2000truncation}
On the other hand, particular attention has been paid during the AMOEBA parametrization to accurately reproduce condensed-phase properties of solvents (and in particular of liquid water). It is then very attractive to combine both models in order to benefit from the best of both worlds getting the small molecule quantum mechanical quality of ANI while keeping the robustness of AMOEBA for condensed phase simulations. This can be achieved by writing the total potential energy of the so-called ANI-2X/AMOEBA hybrid model as
\begin{align}\label{eq:hybrid_potential}
    V_\ttiny{HYB}(P\cup W) &= V_\ttiny{AMOEBA}(P\cup W) + V_\ttiny{ML}(P) - V_\ttiny{AMOEBA}(P)\\
    &= V_\ttiny{AMOEBA}(W) + V_\ttiny{AMOEBA}(P\cap W) + V_\ttiny{ML}(P)\nonumber
\end{align}
where $P$ indicates the solute, $W$ indicates the solvent, $P\cap W$ indicates the solute-solvent interactions and $P\cup W$ indicates the total system. The many-body nature of the polarization energy prevents us from directly computing $V_\ttiny{AMOEBA}(P\cap W)$. To embed the ML potential, we subtract the AMOEBA potential of the isolated solute to the full AMOEBA potential. As indicated in Eq.~\eqref{eq:hybrid_potential}, this is essentially equivalent to using AMOEBA for the solvent-solvent and solvent-solute interactions and the ML model for the solute-solute interactions. The atomic environments that are given to the ML potential therefore only comprise atoms from the solute and should be similar to data present in the training set, thus reducing occurrences of extrapolation. This coupling with AMOEBA allows to simulate atom types not available with MLPs and to include counter ions that are crucial in biology. This also enables the use of the accurate AMOEBA water model while benefiting from the automatic inclusion of long-range effects via AMOEBA's efficient Particle Mesh Ewald periodic boundary conditions.

\subsection{Deep-HP: A Multi-GPU MLP platform within \TinkerHP}

\subsubsection{A General Machine Learning Platform}
New ML architectures are introduced daily and dedicated machine learning libraries \PyTorch, TensorFlow and Keras, have created a large community of developers and users. \citep{pytorchpaper, tensorflowpaper, keraspaper}

Conversely, most of the MD codes (CHARMM, GROMACS, \TinkerHP, ...)\citep{CHARMM,GROMACS}, are often written using compiled languages such as \Fortran\ or \C/\C++. To allow for the simultaneous execution of both \Python-based MLP codes and \TinkerHP\ we implemented an interface that allows for efficient data exchanges between environments while maintaining \TinkerHP\ as the master process which, punctually, calls the MLP code. Identified by \TinkerHP\ as another computational subroutine, the MLP code should be therefore provided as a Python API. We have implemented such functionality using the C Foreign Function Interface (\cffi) for Python which allows for efficient API embedding, within a dynamic library (DLL) to be linked with. Technically, within such a framework we can now call Python frozen codes from C using such \cffi\ embedding feature, thus enabling the use of various MLP codes within \TinkerHP.

In that context, the recent \gpu-accelerated version of \TinkerHP\ \cite{tinkerhpgpu} offers the opportunity to build an overall very efficient hybrid MD/MLP code as both applications are running on the same \gpu\ platform. To do so, we need to design a Python/C interface in a way that avoids any substantial data transfers between \Python\ and \C\ environments. In practice, the \cffi\ module is not natively designed to interface data structures from device memory: its dictionary can only process host addresses on array datatype or scalar data structures. Based on these constraints, our code would be forced to perform two host-device data transfers in order to communicate through \Fortran/\C\ and Python interface. To overcome this issue that would be detrimental to the global performance, we directly send generic memory addresses through the interface as scalar values and use the PyCUDA python module to manually cast these addresses into Tensor type that can actually be used by MLP codes. Fortunately, \PyCUDA\ and \PyTorch\ provide such casting routines. Thus, calling Python codes from \Fortran/\C\ with device data among the calling arguments can be done independently of the size of those arguments.


Furthermore, we built the interface of the MLP code in order to keep \TinkerHP\ model-agnostic. In practice, \TinkerHP\ provides positions and neighbor lists and gets energies and forces in return. Adding a new MLP to the platform then becomes an easy task, especially if it was developed using the \PyTorch\ or \TensorFlow\ libraries. Moreover, we implemented an API within \TorchANI\ which allows to save and reconstruct ANI-like models using JSON, YAML and PKL formats. This allows to directly use models trained with \TorchANI\ with the Deep-HP platform, thus reducing the hassle of transferring a model from the training stage to production simulations.

\subsubsection{Massive Parallelism within \TinkerHP: Scalable Neural Networks Simulations}
Regarding parallelism, \TinkerHP\ uses a three dimensional domain decomposition (DD) scheme. The simulation box is decomposed into a certain number of domains matching the exact number of parallel processes at our disposal so that each process - attached or not to a device - is assigned to a unique domain. Then, each process computes partial forces on the local atoms, communicates the partial data to his spatial neighbors, sums the partial forces and integrates the equations of motions for local atoms at each time-step. The DD method is valid and effective under the assumption that all interactions are short-range and the atomic positions do not move much between two time-steps. The same structure has been used during the development of the accelerated multi-GPU version. \cite{tinkerhpgpu} Naturally, we wanted to preserve this property with the MLP code interface despite the fact that \TorchANI\ is not designed to run on multiple \GPU. Using the DD method from \TinkerHP, we can isolate the local atoms of a domain and its neighbors and send the information to a MLP code instance through the interface for calculation. We also  bypass the implemented neighbor list within \TorchANI, and use the one of \TinkerHP. 
Indeed, we verified that the \TorchANI\ neighbor list algorithm scales as $\mathcal{O}(N^2)$ ($N$ being the number of atoms), both in execution time and memory; which limits its applicability to small systems. For instance, a \num{12000} atoms water box on a Quadro GV100 \gpu\ card supported by \SI{32}{GB} of memory already caused a memory overflow. 
Because \TorchANI\ requires a pair list of indices as a data structure, we adapted the highly \gpu-optimized linked-cell method, thoroughly described in ref.~\cite{tinkerhpgpu}. 
In practice, the list is built by partitioning the box into smaller ones and resort to an adjacency matrix and a filtering process.
Finally, the complexity of the neighbor list generation outperforms the original \TorchANI\ implementation, thus significantly reducing both the computational cost and memory footprint and allowing to handle much larger systems.
For example, systems made of more than \num{100000} atoms are now manageable on a single \SI{32}{GB} GV100 \gpu. On top of that, we also noticed a constant memory allocation from Python (especially when running in parallel) which happens to be detrimental to performance and, on some occasions, can lead to a crash. This issue has been solved by resorting to an upstream bounded buffer reservation which size is proportional to the number of atoms in the system. In the end, Deep-HP is able to perform simulations of several million atoms systems, as illustrated in Figure \ref{fig:perf_1GPU_comp} where we show the scalability of the platform on water boxes up to 7.7 million atoms using up to 68 V100 \GPU.

\subsection{Performance and Scalability Results}

\subsubsection{Benchmark Systems}
We use water boxes of increasing size as benchmark systems as well as some solvated proteins.\cite{tinkerhp,tinkerhpgpu} The solvated proteins and their respective number of atoms, in parentheses, are: DHFR protein (\num{23558}), SARS-CoV2 M$^{pro}$ protein (\num{98500}) and COX protein (\num{174219}). For the water boxes: \num{648} (i.e small), \num{4800} (big), \num{12000} (huge), \num{19200} (globe), \num{96000} (puddle), \num{288000} (pond), \num{864000} (lake), \num{2592000} (bay) and \num{7776000} (sea).
After equilibration, we evaluated the performance on short NVE MD simulations.

\subsubsection{GPU Performances}
\begin{figure*}[!htp]
\centering
{\includegraphics[width=.47\linewidth]{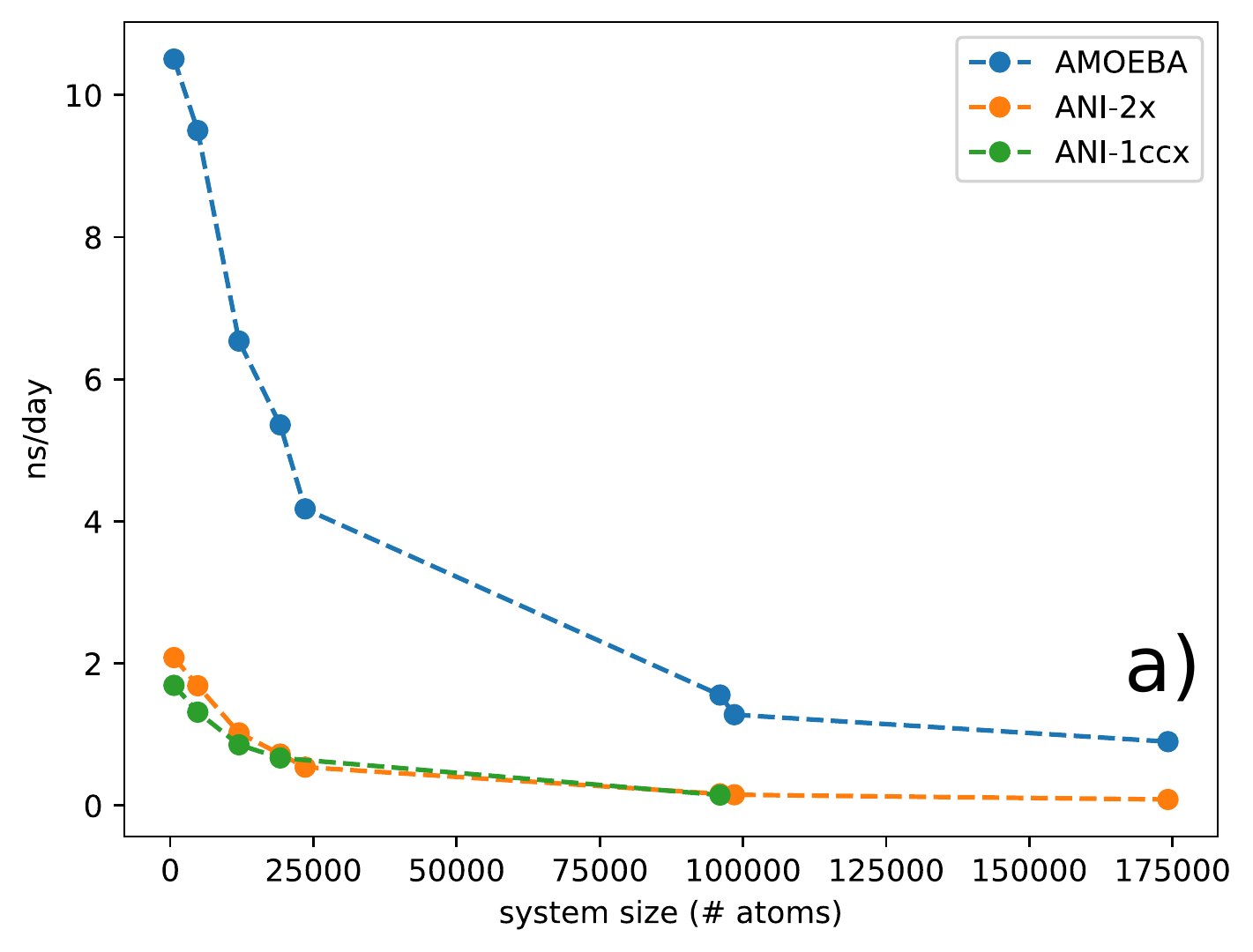}} \hfill
{\includegraphics[width=.47\linewidth]{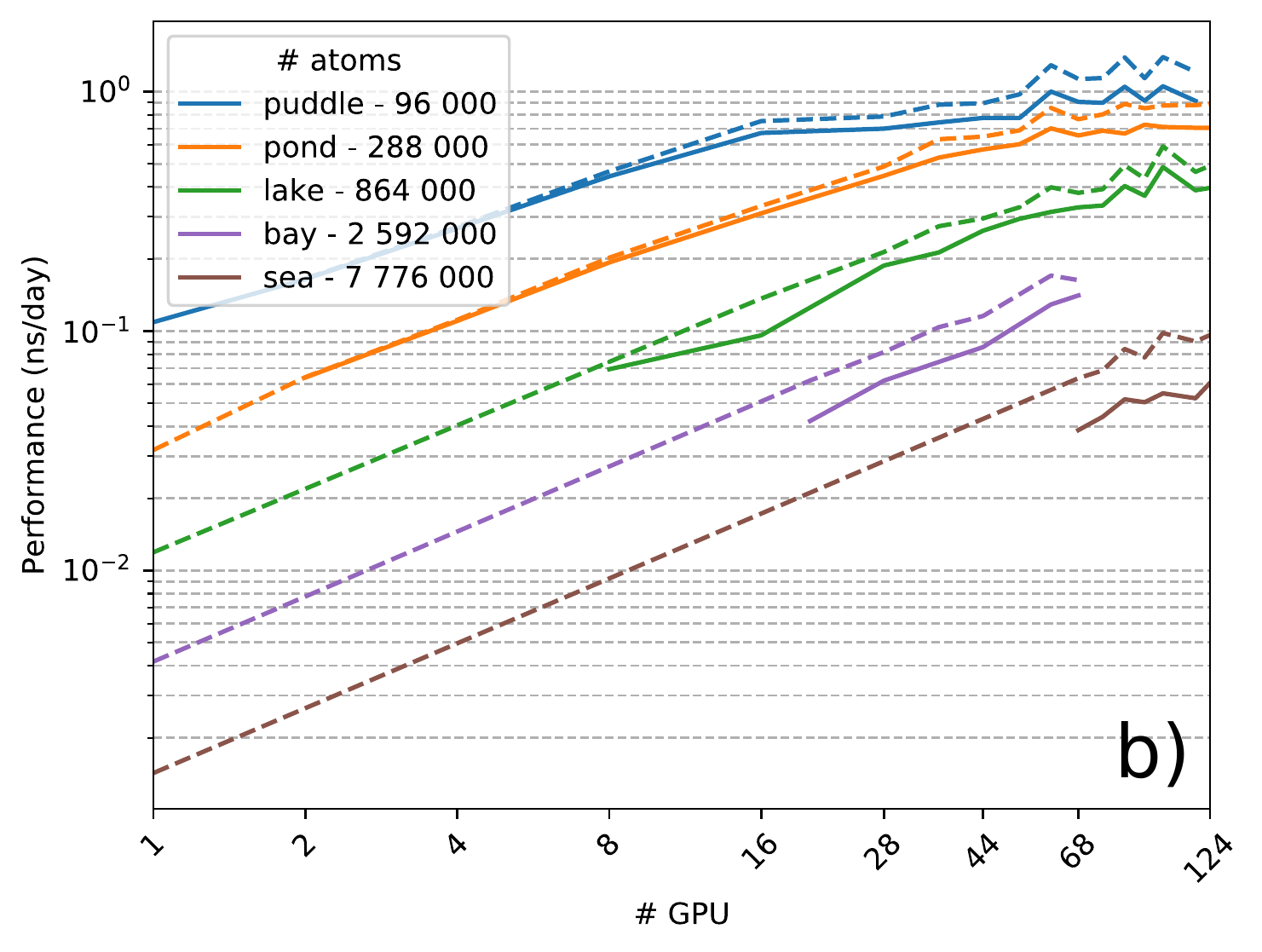}}
\caption{a) Performance comparison between ANI-1ccx(1NN), ANI-2x(1NN) and AMOEBA models in ns per day, over increasing system size, on a single Nvidia Tesla A100. b) Strong scaling logarithmic scale plot of ANI-2x model on benchmark systems. Simulations are performed in the NVE ensemble using a Velocity-Verlet integrator \SI{0.2}{fs} time-step.}
\label{fig:perf_1GPU_comp}
\end{figure*}

To ensure the performance and portability of our platform, we ran tests on different \gpu\ infrastructures such as Tesla V100 nodes of the Jean-Zay supercomputer, the Irène Joliot Curie ATOS Sequana supercomputer V100 partition or a NVIDIA DGX A100 node. In the rest of the text the default device is the Tesla V100 if not mentioned otherwise. For each system, we performed \SI{2.5}{ps} MD simulations with a Verlet integrator using a \SI{0.5}{fs} time-step and average the performance over the complete runs. Figures \ref{fig:perf_1GPU_comp} gather single GPU device performances. 

Before discussing performance results let us introduce three critical concepts: saturation, utilization and peak performance.
Saturation represents the ratio of resources used by the algorithm against the actual resources supplied by the GPU. 
It is closely related to the degree of parallelism expressed within the algorithm and its practical use in the simulation. Given the fact that recent \GPU\ provide and execute several thousands of threads at the same time to run calculations on numerous computational cores, complete saturation is naturally not achieved for small systems.
On the other hand, the device utilization represents the percentage of execution time during which the \gpu\ is active.
As the \gpu\ is driven by the \cpu, its utilization heavily depends on both the \cpu\ speed and the amount of code actually offloaded to the device. It is essential to rely on asynchronous computation and to develop a device-resident application in order to achieve a complete \gpu\ utilization over time.
Finally, peak performance (PP) describes how an algorithm asymptotically harnesses the computational power of the device on which it operates. Increasing this metric implies to maximize arithmetic operations over memory. However, one can only assess device peak performance in terms of floating points operations when both saturation and utilization are maximized. 
With a typical HPC device such as Quadro GV100 which delivers over \SI{15.6}{TFlop/s} in single precision arithmetic ( 4 bytes ), around 69 arithmetic operations can be performed between two consecutive float transactions from global memory, in order to reach the peak performance.
Knowing this, we analyze the \gpu\ peak performance of Deep-HP and \TinkerHP\ AMOEBA, in both separate and hybrid runs, using the reference GV100 Card. Results are depicted in Table \ref{tab:peak_perfs}. We can see the influence of device saturation on peak performance while running pure ML models, from the under-saturated DHFR system to the over-saturated COX one. MLPs manage to reach excellent peak performance on \gpu\ platforms due to the large amount of calculations induced by the numerous matrix-vector products involved.
For AMOEBA, on the other hand, the relatively tiny increase of peak performance for both systems - second column of Table \ref{tab:peak_perfs} - denotes an excellent saturation and utilization of the device, regardless the size. The overall peak, however, reaches a lower \SI{10.52}{\%}, which is still satisfactory given the complexity of the algorithm involved in the PFF calculation.

\begin{table}[]
    \centering
    \begin{tabular}{|c|lll|}
    \hline
    System/Model   & ANI & AMOEBA & Hybrid\\ \hline
    DHFR     & 19.42  & 9.08   & 5.16\\
    COX      & 28.13  & 10.52  & n/a \\ \hline
    \end{tabular}
    \caption{Global Peak performance in percentage(\%) assessed over a 50 femtoseconds MD trajectory. The Quadro GV100 was chosen to be the reference device. }
    \label{tab:peak_perfs}
\end{table}

To study the complexity of the algorithm, we ran the benchmark systems on a single DGX A100 with two ANI models and compared the performance against the AMOEBA force field (see Figure \ref{fig:perf_1GPU_comp}a). The ANI-1ccx simulations are performed on water boxes ranging from \num{648} to \num{96000} atoms. For ANI-2X we also considered three solvated proteins: DHFR, SARS-CoV2 M$^{pro}$ and COX. Furthermore, for these tests, we performed inference using only one instance from the ensemble of eight neural network predictors of the ANI models.
On water boxes, ANI-1ccx is found to be between 2 and 7\% faster than ANI-2X due to the models intrinsic complexities. Figure \ref{fig:perf_1GPU_comp} a) shows the performance of both ANI-2X and AMOEBA. On the \num{648} and \num{4800} atoms systems, AMOEBA is \num{1.85} and \num{2.20} times faster than ANI respectively. On the first four water systems the ratio grows as $\mathcal{O}(N)$ with respect to the number of atoms $N$, with a Pearson coefficient equal to \num{0.995}. On the protein systems the ratio still grows linearly but with a smaller slope: roughly a factor 2 is preserved.
\begin{table*}[htp!]
\newcommand{\na}{\textsf{n/a}}
\begin{center}
\begin{tabular}{|m{9em}|*{10}{c}|} 
    \hline
    \rowcolor{Gray}
    Systems (Number of atoms)/ Number of GPU devices & 1 & 4 & 8 & 16 & 28 & 44 & 68 & 84 & 100 & 124 \\[6pt] \hline
    \rowcolor{Gray} & \multicolumn{10}{c}{ GPU V100 } \\[3pt] \hline
    Puddle(96000)    & \num{0.11} & \num{0.27} & \num{0.44} & \num{0.67} & \num{0.70} & \num{0.78} & \num{0.91} & \num{1.05} & \num{1.05} & \num{1.05} \\
    Pond(288000)      & \na & \num{0.11} & \num{0.19} & \num{0.31} & \num{0.46} & \num{0.57} & \num{0.66} & \num{0.67} & \num{0.71} & \num{0.71} \\
    Lake(864000)       & \na & \na & \num{0.07} & \num{0.10} & \num{0.19} & \num{0.26} & \num{0.33} & \num{0.40} & \num{0.48} & \num{0.40}  \\
    Bay(2592000)        & \na & $\ldots$ & \na & \num{0.04} & \num{0.06} & \num{0.09} & \num{0.14} & \na & \na & \na \\
    Sea(7776000)        & \na &  & $\ldots$ & $\ldots$ &  & \na & \num{0.04} & \num{0.05} & \num{0.06} & \num{0.06}  \\ \hline
    \rowcolor{Gray} & \multicolumn{10}{c}{ GPU A100 } \\[3pt] \hline
    Puddle(96000)     & \num{0.16} & \num{0.41} & \num{0.63} & \na &  &  & $\ldots$ &  &  & \na \\
    Pond(288000)       & \na & \num{0.16} & \num{0.26} & \na &  &  & $\ldots$ &  &  & \na \\
    Lake(864000)       & \na & \na & \num{0.11}& \na &  &  & $\ldots$ &  &  & \na \\
    \rowcolor{Gray} & \multicolumn{10}{c}{ Theoretical performance (V100)} \\[3pt] \hline
    Puddle(96000)     & \num{0.11} & \num{0.27} & \num{0.46} & \num{0.75} & \num{0.79} & \num{0.90} & \num{1.14} & \num{1.39} & \num{1.40} & \num{1.40} \\
    Pond(288000)       & \num{0.03} & \num{0.11} & \num{0.20} & \num{0.33} & \num{0.49} & \num{0.65} & \num{0.77} & \num{0.89} & \num{0.88} & \num{0.89} \\
    Lake(864000)       & \num{0.01} & \num{0.02} & \num{0.07} & \num{0.14} & \num{0.21} & \num{0.30} & \num{0.38} & \num{0.49} & \num{0.59} & \num{0.49} \\
    Bay(2592000)        & \num{0.004} & \num{0.007} & \num{0.02} & \num{0.06} & \num{0.08} & \num{0.12} & \num{0.16} & \na & \na & \na \\
    Sea(7776000)        & \num{0.001} & \num{0.003} & \num{0.005} & \num{0.009} & \num{0.02} & \num{0.03} & \num{0.06} & \num{0.08} & \num{0.10} & \num{0.10} \\
    \bottomrule
\end{tabular}
\caption{\label{tab:scalability}.   Performance of the ANI-2x neural network in Deep-HP in term of molecular dynamics simulation production (ns/day) for selected water boxes of increasing sizes using Nvidia V100 and A100 GPU cards. \na\textit{: not available}}
\end{center}
\end{table*}

To further analyze the computational bottleneck of HDNNP models, we evaluated the contribution of each of the model's constituents to the overall execution time (Figure 1 SI). For small systems more than 40$\%$ of the cost is due to the gradients and AEV computations. The \TinkerHP\ neighbor list is less than 5$\%$ of the cost, demonstrating the performance of the implementation. For larger systems, the computational cost is largely dominated by the gradients computation (i.e. more than 50$\%$). Thus, ML potentials computational performances are now mainly limited by back-propagation and not by the environment vector (the latter mainly being the memory bottleneck). Accelerating the gradients estimation will therefore be of the utmost importance for future implementations. Deep-HP also provides a keyword to automatically use mixed precision within \PyTorch. The automatic mixed precision is using a combination of half and single precision operations without a severe loss on the model's accuracy. 

\subsubsection{Multi-\gpu\ Performance and Scalability of ANI Models within \TinkerHP}
In the following, we assess and discuss the multi-node performance of Deep-HP. The Jean Zay HPE SGI \num{8600} \gpu\ system holds numerous computing nodes accelerated by \num{4} interconnected Tesla V100 devices each. Ideally, a parallel algorithm associated to a certain amount of resources (N processors for instance), whose load is equally distributed across all resources, will exactly perform N times faster. Experimentally, an intermediate step, occupied with communications, affects the performance to a varying degree depending on size and pattern of these communications in comparison with the amount of calculations. When the number of allocated resources increases, global synchronizations induced by collective communications significantly slow down the parallel execution and, therefore, impact the asymptotic behavior of the strong scalability. Communication patterns and speed are subsequently the principal obstacles to achieve an ideal scaling.
In our case, the domain decomposition method coupled with ANI offers an up-bounded communication pattern, which allows to use several nodes without enduring severe performance loss too quickly, as it is the case with multi-node PFF on \GPU. \cite{tinkerhpgpu}
As displayed in Figure \ref{fig:perf_1GPU_comp} b), we are able to scale up to \num{11} nodes ( \num{44} devices ) for a \num{864000} atoms water box, before suffering from communication overheads and insufficient load. On the other hand, note that an accurate estimation of the gradients for each atom requires a complete knowledge of its surrounding environment up to a predetermined distance.
The current implementation is however not optimal for a large number of processes and the performance starts to cap when half the minimum length of a domain equals the cutoff distance of the atomic environments. This is due to some redundancy between processes for the calculation of AEVs and energies of atoms from neighbouring domains. To illustrate this effect, we made an estimation of the performance in the case of no computational redundancy and plot it for every test case in dashed lines within Figure \ref{fig:perf_1GPU_comp} b). As anticipated, dealing with this effect can offer a significant 40\% boost in the parallel run as is observed for the sea water box. Thus, future implementations should address this issue in order to maximize multi-nodes performance. The test machines we used were also not optimal and do not provide fast interconnect between nodes. The observed A100 50\% boost coupled to improved nodes interconnections will certainly be extremely beneficial to Deep-HP (we could not get access to a large recent A100 cluster and were limited to a single DGX-A100 node). Nevertheless, the current implementation can already be considered as a game changer for ANI/ANI-2x DNNs simulations as the use of several GPUs already provides the capability to produce ns/day molecular dynamics simulations on hundreds of thousands atom systems (see detailed benchmarks on Table \ref{tab:scalability})




\begin{table*}[htp!]
\begin{center}
\noindent\makebox[\textwidth]{
\begin{tabular}{ccccccc} 
    \hline
    \rowcolor{Gray}
    {splits} & {0.2} & {0.25/1} & {0.25/2} & {0.25/2/4} & {0.25/2/6} \\[2pt] \hline
    Benzene$^a$  & 1.0  & 4.74  & 8.42 & 14.51 & 18.17 \\
    Water$^a$  & 1.0  & 4.39  & 8.07 & 12.58 & - \\
    Benzene$^b$  & 1.21  & 5.74  & 10.20 & 17.57 & 22.00 \\
    Water$^b$  & 2.03  & 8.92  & 16.40 & 25.57 & - \\
    Integrator-type  & V  & R  & R  & R1(HMR) & R1(HMR)  \\ \bottomrule
\end{tabular}
}
\caption{\label{tab:speedint}. Relative speedup of hybrid models with RESPA (R) and RESPA1 (R1) Integrators calculated with respect to: $^a$ hybrid model Velocity-Verlet (V) 0.2 fs time step; $^b$ ANI only with Velocity-Verlet (V) 0.2 fs time step.}
\end{center}
\end{table*}

\subsubsection{Accelerating Hybrid Simulations: Multi-timestep Integrators (RESPA/RESPA1) and Reweighting Strategies}
\subsubsection{Multi-timestep Integrators (RESPA/RESPA1)}
As fast as the ANI model can be compared to Density Functional Theory (~10**6 factor speedup), ANI remains far more computationally demanding than polarizable force fields (see SI, Tables 2 and 3) and the stiff intramolecular interactions reproduced by the MLP limits the integration time-step to "ab initio" 0.2-0.3 fs values, thus making the study of large proteins on long biological timescales a daunting task. One way to speed up MD is to use larger time steps through multi-time-stepping (MTS) methods thanks to an hybrid model. As discussed in Section 2.5, we decided to introduce the ANI-2X/AMOEBA model that is coupling a very accurate MLP for small molecules (ANI) to a PFFS designed to produce accurate condensed phase simulations of solvated proteins (AMOEBA). 
Typical MTS schemes exploit the separability of the potential energy into a computationally expensive, slowly varying part and a cheap, quickly varying part, and use a specific integration scheme, RESPA \cite{respa}, that allows for less frequent evaluations of the expensive part. In particular in the context of the AMOEBA PFF, \TinkerHP\ uses either a bonded/non-bonded splitting or a three-stage separation between bonded, short-range non-bonded and long-range non-bonded interactions\cite{mts_int_PFF} (denoted as RESPA1 in the rest of the text). In both cases, temperature control is made through a BAOAB discretization of a Langevin equation\cite{leimkuhler2013robust}. In this context, the bonded forces are integrated using a small 0.2-0.3 fs time-step and the outermost time-step can be taken as \SI{2}{fs} or \SI{6}{fs} depending on the splitting. These can be further pushed by using Hydrogen Mass Repartitioning (HMR)\citep{split_PFF, mts_int_PFF}.
These integration schemes extend the applicability of PFFs to longer time-scale reducing the gap with classical FFs, as demonstrated with recent simulations of tens of $\mu s$ of the SARS-CoV2 M$^{pro}$ protease\citep{proteaseadaptive}.

Even though MLPs are much less expensive than \textit{ab initio} calculations, the most common MLPs with feed-forward neural networks remain more computationaly demanding than FFs, even polarizable ones (see SI, Table 2). To reduce this gap, towards simulating large biological systems, we combined our hybrid ANI-2X/AMOEBA model to MTS integrators using the RESPA scheme. 
We assume that AMOEBA is a good approximation of the ML potential for the isolated solute so that their energy difference $\Delta V_\ttiny{ML}(P)=V_\ttiny{ML}(P) - V_\ttiny{AMOEBA}(P)$ should produce small forces that can be integrated using a larger time-step. This is done in the same spirit as Liberatore et al. \cite{liberatore2018versatile} that studied such integration scheme in the context of accelerating ab initio molecular dynamics. We thus associate this difference with the non-bonded part of the AMOEBA model and end up with the following separation:
\begin{align}\label{eq:respa}
    V^{fast}_\ttiny{HYB}(P\cup W) &= V^{bond}_\ttiny{AMOEBA}(P\cup W) \\
    V^{slow}_\ttiny{HYB}(P\cup W) &= \Delta V_\ttiny{ML}(P) + V^{nonbond}_\ttiny{AMOEBA}(P\cup W)
\end{align}
where $ V^{fast}_\ttiny{HYB}$ is evaluated every inner time-step and $V^{slow}_\ttiny{HYB}$ every outer one. 
In the RESPA1 framework, the potential energy difference $\Delta V_\ttiny{ML}(P)$ is associated with the long-range interactions and evaluated at the outermost time-step.

To assess the accuracy of each integrator we computed the solvation free energy of two solute with the hybrid model described above: the benzene molecule solvated in a cubic box of \num{996} water molecules with a \SI{31}{\angstrom} edge and a water molecule in a cubic box of \num{3999} other water molecules with a \SI{49}{\angstrom} edge. For each of these systems and integrators, we computed their solvation free energy by running 21 independent trajectories of \SI{2}{ns} and \SI{5}{ns} where the ligand is progressively decoupled from its water environment, first by annihilating its permanent multipoles and polarizabilities and then by scaling the associated van der Waals interactions (while using a softcore). The trajectories were run in the NPT ensemble at \SI{300}{K} and 1 atmosphere using a Berendsen barostat and either a Bussi thermostat\cite{bussi2007canonical} (when Velocity Verlet is used) or a Langevin one for the MTS simulations as mentioned previously. The free energy differences where then computed using the BAR method\cite{bennett1976efficient, sampling_jerome_lucie}. Results were compared with a reference Velocity-Verlet integrator using a \SI{0.2}{fs} time-step. The AMOEBA bonded forces were always evaluated every \SI{0.25}{fs}. In the case of a bonded/non-bonded split, the non-bonded forces were evaluated either every 1 or \SI{2}{fs}, and in the case where the non-bonded forces are further split between short-range and long-range ones, the short-range non-bonded forces were evaluated every \SI{2}{fs} and the long-range ones either every \SI{4}{fs} or \SI{6}{fs}. As explained above, the MLP forces are always computed at the outermost time-step.
Table \ref{tab:speedint} shows the speedup of our hybrid model with various MTS setups compared to reference Velocity Verlet ANI-2X/AMOEBA simulation with a \SI{0.2}{fs} time-step and Velocity Verlet ANI simulations with a \SI{0.2}{fs}. In practice, speedups are system-dependant, but RESPA techniques always lead to a minimal acceleration of an order of magnitude with the tighter accuracy integration scheme (RESPA split) for a ANI solute in an polarizable AMOEBA solvent and compared to an ANI (Verlet 0.2fs) simulation setup. Concerning the accuracy, results are displayed in SI (see Table 1). RESPA1 approaches, despite being operational, appear more sensitive to the system and do not always lead to the desired result in term of free energies and should be restricted to simple simulation purposes. Therefore, the tighter RESPA (0.25/1 and 0.25/2) integrators are found to be good compromises between accuracy and computational gain.
These integrators thus extend the applicability of machine learning-driven molecular dynamics to larger biologically-relevant systems and to longer-time-scale simulations. In practice, the resulting performance gain helps to reduce the computational gap between ANI and AMOEBA that is initially about more than a factor 30 (see SI, Table 2).
\subsubsection{Accelerating Hybrid Simulations: an Alternative Reweighting Strategy}
Concerning the proposed multi-timestep approach, it is important to note that since we assume that AMOEBA is a good approximation of the ML potential for the isolated solute, the present acceleration strategy is not possible when this condition is not fullfield. In practice, it could happen in the event of an intramolecular reaction within the DNN solute. Indeed, ANI-2X being a reactive potential, it is sometime able to produce intramolecular proton transfers in some specific cases, i.e. when donor and acceptor functional groups are present. On the opposite, AMOEBA is non-reactive force field that will always stay in its initial electronic state. Therefore an intra-ligand chemical reaction would desynchronize the two potentials and therefore stop the simulation. In practice, it is not an issue since such ANI-2X extra-feature introduce additional useful interpretative information about the possible ligand states and does not block the evaluation of free energies. In the rare case of such an event, it is always possible to use a re-analyse approach and to produce the BAR simulation windows thanks to fast RESPA AMOEBA, non-reactive, trajectories. Then one can re-analyse the saved AMOEBA snapshots, by computing the corresponding ANI-2X/AMOEBA energies to correct the AMOEBA free energy evaluation using a rigorous BAR reweighting \cite{zhang2012polarizable, sampling_jerome_lucie} (details can be found in SI, see section 2.2). Such an alternative approach conserves the advantage of speed since the computation of the costly DNN gradients are avoided. It can also be beneficial in rare cases, where AMOEBA and ANI-2X potential differ too much or when one would like to benefit from extra sampling capabilities thanks to the more affordable AMOEBA MD computational performances. 

\begin{figure*}[!htp]
    \centering
    \subfloat{\label{a}\includegraphics[width=.5\linewidth]{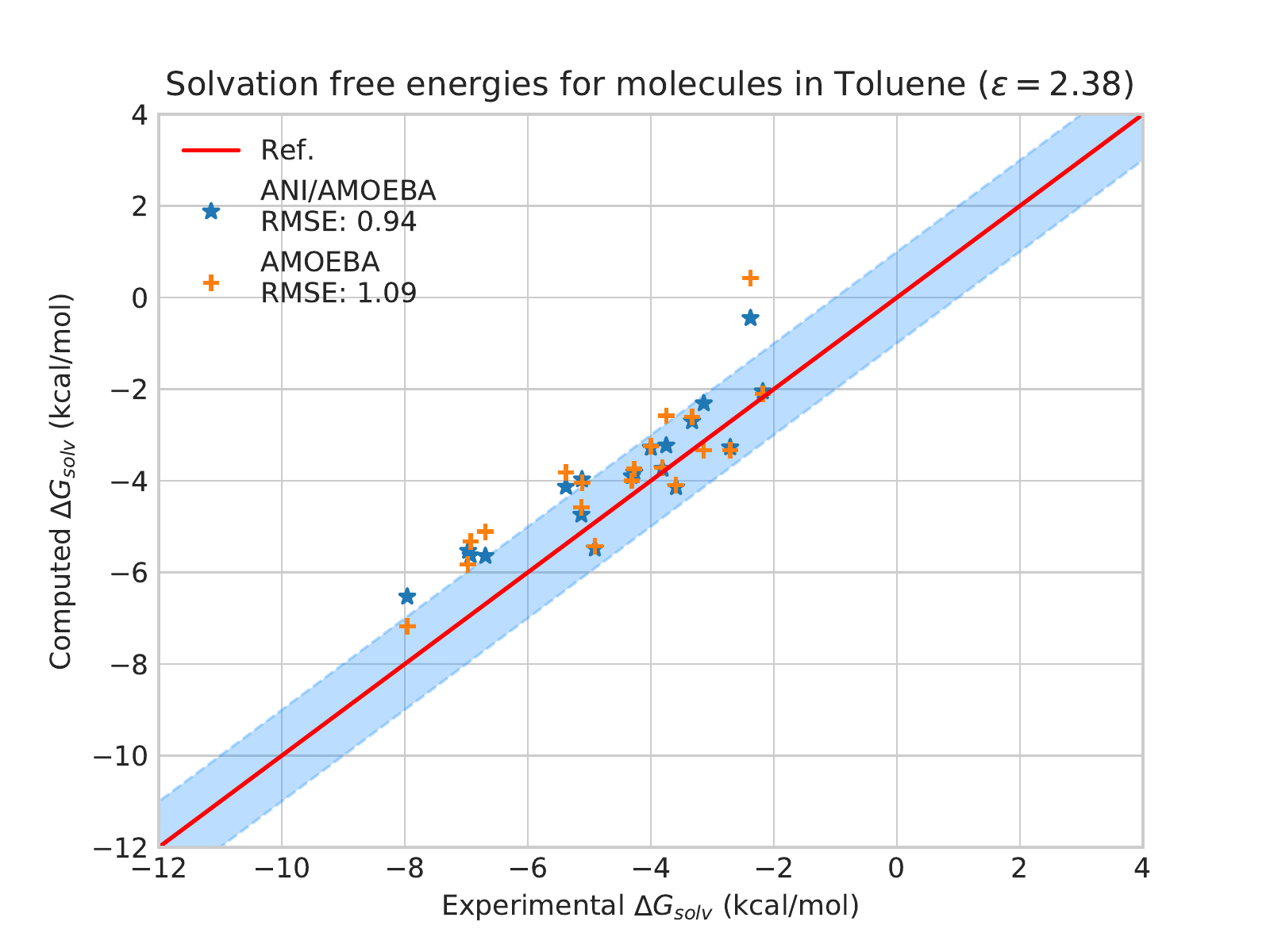}}\hfill
    \subfloat{\label{b}\includegraphics[width=.5\linewidth]{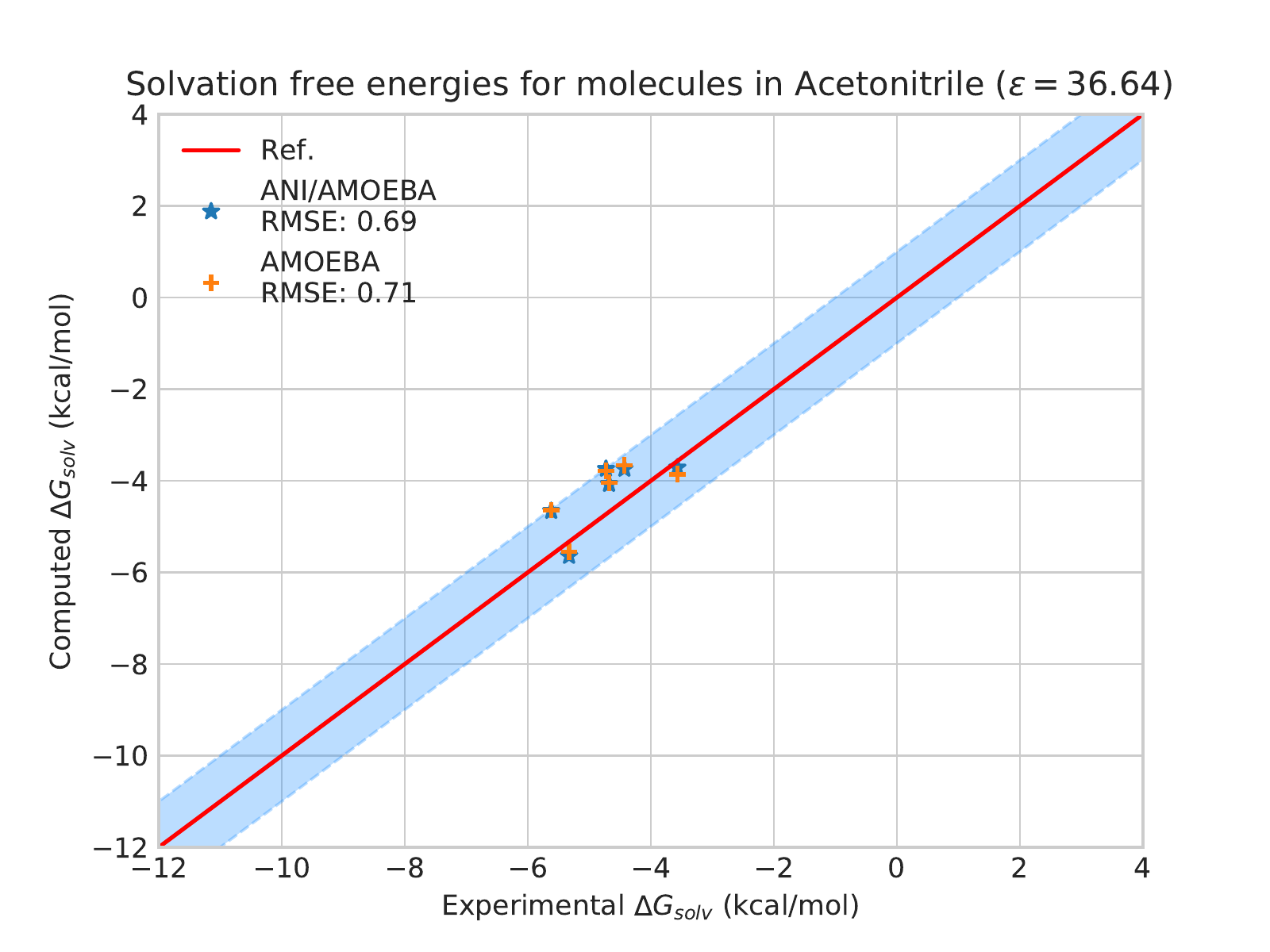}}\par

    \subfloat{\label{c}\includegraphics[width=.5\linewidth]{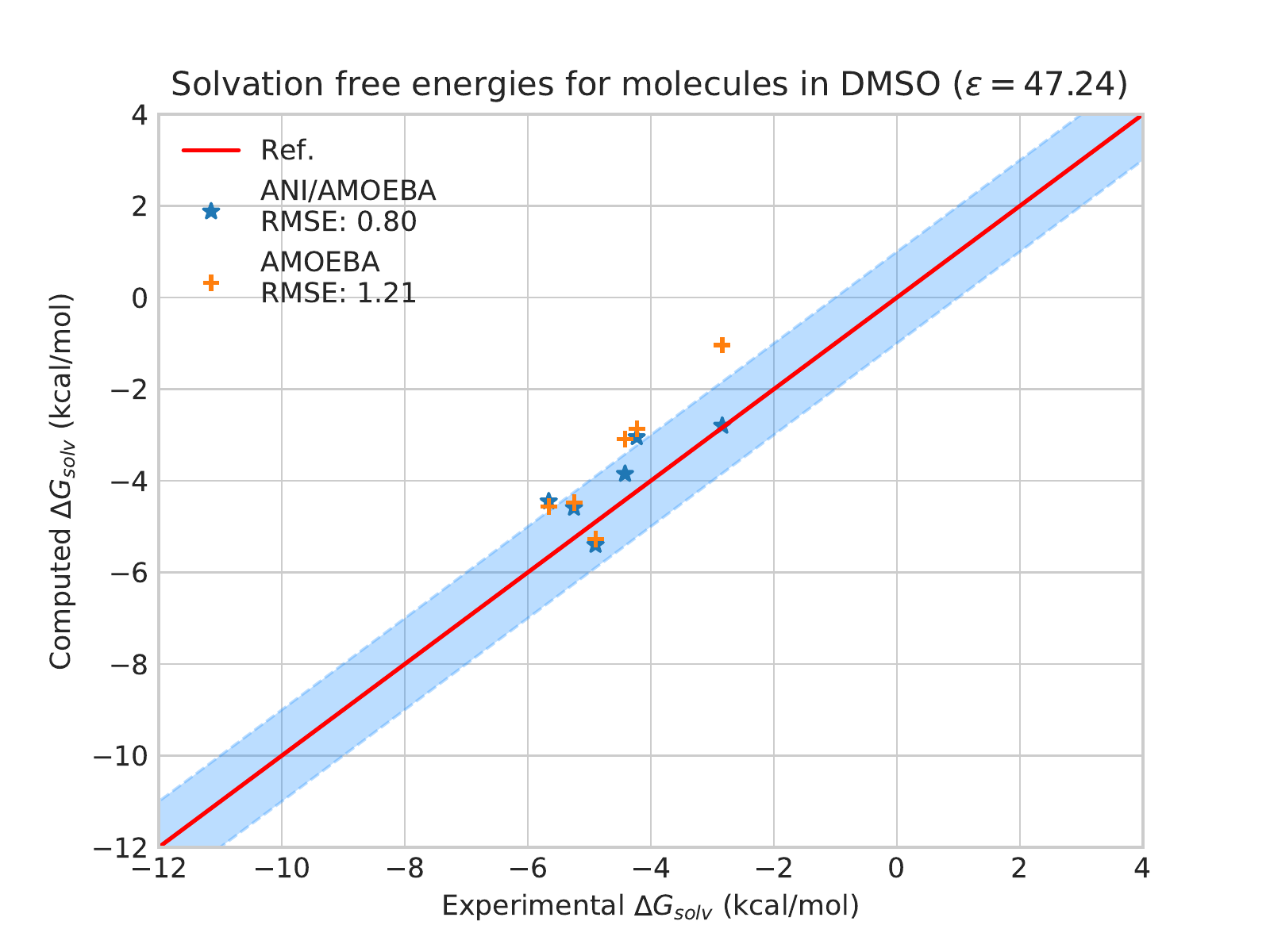}}\hfill
    \subfloat{\label{d}\includegraphics[width=.5\linewidth]{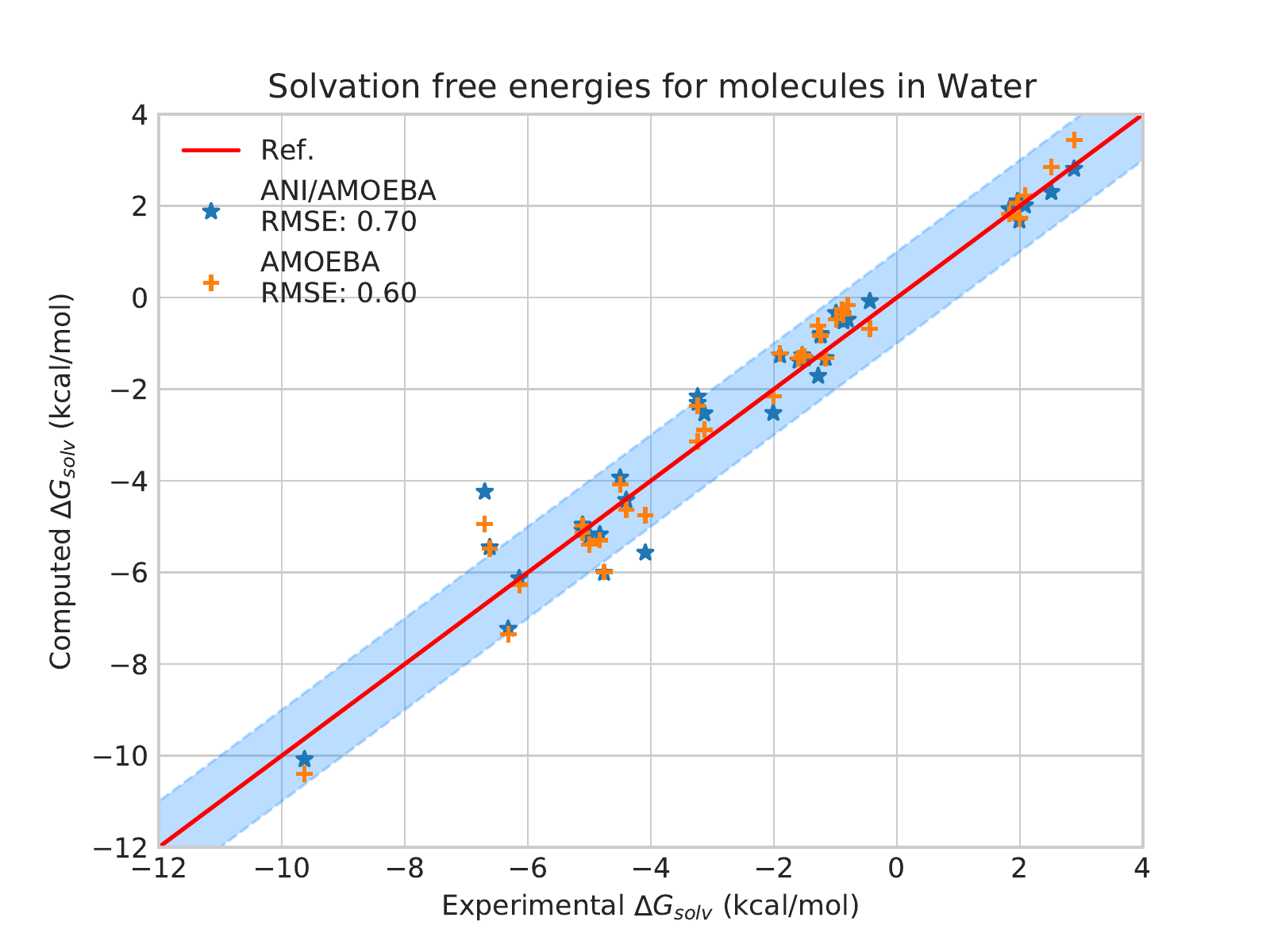}}\par
    \caption{Solvation free energies of molecules in different solvents computed with AMOEBA (orange) from refs \citep{essex_free_energies, poltype1} versus hybrid model ANI-2X/AMOEBA (blue) and experiment (red). The blue domain correspond to the so-called chemical accuracy: error of 1 kcal/mol w.r.t experiment.}
    \label{fig:solv_freeenergies}
\end{figure*}

\section{Results}

\subsection{Solvatation Free Energies}
\subsubsection{Computational Details}
To assess further the performance of the ANI-2X/AMOEBA hybrid model, we extended our solvation free energies tests to a variety of small molecules at non-aqueous and aqueous conditions following references \cite{essex_free_energies, poltype1}. The considered solvents (with their dielectric permittivity) are: Toluene ($\epsilon=2.38$), Acetonitrile ($\epsilon=36.64$),  DMSO  ($\epsilon=47.24$) and water ($\epsilon=77.16$). The description of the solutes can be found in SI. 
 \\
We withdrew molecules from the dataset that have chemical elements that were not available in ANI-2x. This led us to a total of 39 molecules solvated in water (taken from reference\cite{poltype2}), 20 molecules solvated in toluene, 6 in acetonitrile and 6 in DMSO (taken from Essex et al).\citep{essex_free_energies}. 
All the systems were prepared following standard equilibration protocol: after a geometry optimization, they were progressively heated up to \SI{300}{K} in NVT and then equilibrated for \SI{1}{ns} in the NPT ensemble at the same temperature and 1 atmosphere. In all cases, we used the most simple multiple time-step integrator presented above with a \SI{0.25}{fs} time-step for bonded terms and \SI{1}{fs} for the outermost one. The Bussi thermostat and the Berendsen barostat were used.
The van der Waals interactions cutoff was chosen at \SI{12}{\angstrom} and the electrostatic interactions were handled with the Smooth Particle Mesh Ewald method\cite{SPME} with a \SI{7}{\angstrom} real space cutoff and default Tinker-HP grid size. 
We used the same scheme as before to decouple the systems from their environment with 21 independent windows of \SI{2}{ns}. For solvation free energies in water we also pushed the ANI-2X/AMOEBA simulation windows up to 5 ns. Indeed, water have been intensively studied as they are an essential tool in drug design and allow to test for the validity of various computational methods and models\cite{D1SC05892D, jorgensen_covid}. The results are compared with experimental data and with the AMOEBA ones. Note that we use both ANI-2X and AMOEBA standard available parametrization without any attempt of further improvement.
\subsubsection{Results and Discussion}
The experimental, AMOEBA and ANI-2X/AMOEBA solvation free energies data are provided in SI (see Tables S1–S4). In the less polar solvent, Toluene ($\epsilon=2.38$), AMOEBA results appear very similar to the hybrid ANI-2X/AMOEBA ones, with a respective RMSE of \SI{1.06}{kcal/mol} vs \SI{1.09}{kcal/mol}. However, this is mainly due to 2 outliers that have an error greater than \SI{1.5}{kcal/mol} with experiment, especially the methylamine for which AMOEBA exhibits a \SI{0.23}{kcal/mol} error while ANI-2X/AMOEBA culminates at \SI{2.60}{kcal/mol}. By removing the methylamine compound, ANI-2X/AMOEBA RMSE accuracy slightly outperforms AMOEBA as the RMSE becomes respectively \SI{0.93}{kcal/mol} vs \SI{1.09}{kcal/mol}. In Acetonitrile, ANI-2X/AMOEBA is close by \SI{0.02}{kcal/mol} with AMOEBA. Strikingly, in the DMSO, ANI-2X/AMOEBA performs much better than AMOEBA, with a respective RMSE of \SI{0.80}{kcal/mol} vs \SI{1.21}{kcal/mol}.
In water, the hybrid model performs slighlty worse than AMOEBA with a RMSE of \SI{0.86}{kcal/mol} vs \SI{0.6}{kcal/mol} for AMOEBA. However, this is again due to two "amine" outliers (methylamine and dimethylamine). Removing them leads to a RMSE decrease to \SI{0.76}{kcal/mol} for the hybrid model. For water, we decided further the simulation time in order to match the 5ns windows used initially for AMOEBA. With longer windows, the ANI-2X/AMOEBA RMSE stabilizes to \SI{0.70}{kcal/mol}, being roughly identical (within the statistical error) to AMOEBA. This is really satisfactory knowing the known accuracy of the AMOEBA model for such systems.
Overall, the hybrid ANI-2X/AMOEBA approach benefits from the inclusion of the neural network and appears to be more accurate than AMOEBA for three out of the four studied solvents: Toluene, Acetonitrile and DMSO. In contrast, AMOEBA is slightly more accurate in the most polar solvent: ie. water. This result is not surprising as the AMOEBA water model is well-known for its accuracy and capabilities to reproduce numerous water-related experimental data\cite{ren2003polarizable}. However, our results confirm the ANI-2X/AMOEBA robustness which appears surprisingly good in such a polar solvent like water once long-range and many-body effects are present.
In contrast with the results obtained by Lahey and Rowley \cite{Rowley_binding} that showed the difficulties of the ANI-2x potential for modeling charged systems within a hybrid embedding approach with non-polarizable force fields we observed accurate results even for charged systems. This is due to a combination of factors linked to many-body and long-range effects and to solvation. Indeed, in the ANI-2X/AMOEBA framework, the charged ligands are embedded in a flexible polarizable solvent that can adapt it dipolar moment to its micro-environment net charges (see references \cite{D1SC00145K,interfacial_dina}for discussions), providing extra flexibility for the hybrid polarizable embedding approach. For example, the hybrid approach yields good results for nitromethane, which is globally neutral but still bears two charged groups. In this solvation study, the RESPA acceleration strategy has been shown to be particularly effective. The main issue observed in our new ANI-2X/AMOEBA approach is clearly linked to some present identified limitations of the ANI-2X potential, such as the amine groups. This should be improved in a near future. \\
In the next section, we go a step further in terms of complexity and report the hybrid model performance on 14 challenging host-guest systems taken from the SAMPL competitions.\cite{sampl_challenge} 

\subsection{Host-guest Binding Free Energies: SAMPL Challenges}

\subsubsection{Computational details}

We considered the absolute binding free energy values of \num{13} guests from the 14 SAMPL4 CB[7]-guest challenge\cite{sampl_hostbind}. We will consider separately the C5 compound that was previously shown \cite{sampl_hostbind} to be a specific outlayer case. We completed the study adding a fourteen complex, a G9 guest taken from the SAMPL6 cucurbit[8]uril host–guest challenge. Free energies were calculated with the hybrid ANI-2X/AMOEBA model as the difference between the free energy of decoupling the ligands within the host and in solution. The optimized structures and parameters for the AMOEBA FF were taken from literature. \cite{sampl_hostbind, tinkeropenmm_freeenergies, sampl_AMOEBA,laury2018absolute} Again, in order to evaluate the impact of the ANI-2X contributions, no AMOEBA specific parametrization has been performed. These ligands are challenging as they are mostly charged, flexible and large, usually leading to difficulties in the prediction of binding free energies.\cite{sampl_hostbind}
The same protocol (\SI{2}{ns} windows) as before was used except that the RESPA outer time-step was changed from \SI{1}{fs} to \SI{2}{fs} which still give a satisfactory accuracy as can be seen in Table 1 of SI. We also provide the free energy values for extended simulations with \SI{5}{ns} windows in order to explore the accuracy convergence.
\subsubsection{Results and Discussion}
\begin{figure*}[!htp]
    \centering
    \includegraphics[scale=.7]{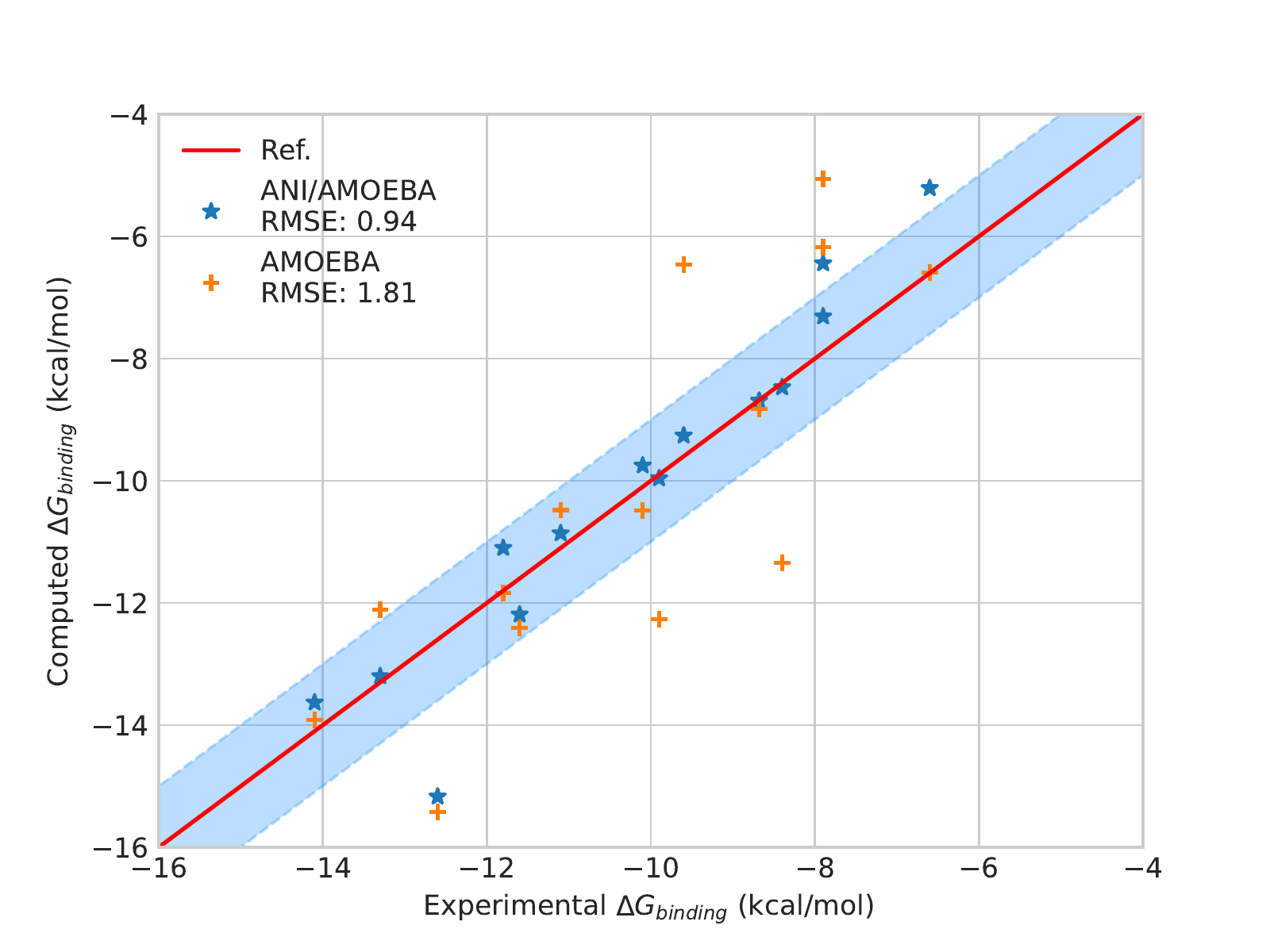}
    \caption{Binding free energies of host-guest systems of the SAMPL4 and SAMPL8 blind challenges with AMOEBA (orange) from refs \citep{sampl_hostbind} versus hybrid model ANI-2X/AMOEBA (blue) and experimental (red). The blue domain correspond to the so-called chemical accuracy: error of \SI{1}{kcal/mol} w.r.t experiment.}
    \label{fig:binding_freeenergies}
\end{figure*}
The binding free energies of the host-guest systems are depicted in Figure \ref{fig:binding_freeenergies} and in SI (see Table S6). Let's focus first on the accuracy of the ANI-2X/AMOEBA prediction. Overall, the hybrid potential results outperform the available AMOEBA data reaching the 1 kcal/mol average error w.r.t experiment on the testset (i.e. RMSE of 0.94 kcal/mol for ANI-2X versus 1.81 kcal/mol for AMOEBA, see Figure 3 ). It is important to note that in this very challenging testset, most of the ligands are charged and encompass a net charge of 1 or 2. Only a single neutral host-guest system extracted from the SAMPL6 challenge is present. The hybrid DNN/PFF performance is very satisfactory as ANI-2X was not designed for modeling ionic species as stated above. As for solvation free energies, the combination of the ANI-2X ligands with the polarizable AMOEBA solvent, host and long-range effects appears to be a powerful tool. ANI-2X/AMOEBA exhibits a larger error than AMOEBA for only the C13, C8 and C3 guest ligands. For C13, predictions are both within 0.5 kcal/mol from experiement. For C8, ANI-2X/AMOEBA stays also within 1 kcal/mol of error (0.7 kcal/mol). C8 has been shown to be associated to high enthalpy changes throughout binding\cite{sampl_hostbind} and such change can be traced back to some gains in term of H-bond interactions from solution to the host-guest complex. It suggests that improvements of ANI-2X towards improved H-bond treatment could be beneficial. This is consistent with our findings on the solvation free energies where AMOEBA performs slightly better than ANI-2X. Concerning C3, the case is more complex and we review our results below in the same section in link with the discussion on integrators' performances. Only, two compound predictions did not reach chemical accuracy: C9 and C10. However, in these cases, the initial AMOEBA error are improved (divided by 2 for C10) using ANI-2X/AMOEBA confirming the higher accuracy of the hybrid model. This result could be associated to slow sampling convergence as noticed by Ren et al. \cite{sampl_hostbind} 
It is worth reporting that on the last compound, i.e. the neutral SAMPL8 host-guest system, the ANI-2X/AMOEBA results almost exactly match experiment (see SI, Table S5). Finally, we also present in SI (Table S6-2), the results for the C5 compound that was removed from the testset. These results confirm the initial assessment by Ren et al.\cite{sampl_hostbind} and would require further investigation (protonation states, binding modes, sampling time etc...) going beyond the scope of the present work.\\
Looking in details at the free energy acceleration strategy, we were overall able to use a RESPA approach on 12 of the 15 (14 + C5) tested ligands. The integrator was not stable enough for the C2, C3 and C4 compounds (see Figure 3 and Table S6-1, SI). This is due to different reasons. First, C2 and C4 exhibited notably higher differences between the ANI-2X and AMOEBA potentials compared to other ligands. This can be easily understood when considering that C2 and C4 are actually associated to the two largest AMOEBA dataset deviations from the experimental reference values (errors of 3.14 and 2.94 kcal/mol, for C2 and C4 respectively). Since our initial choice was to not perform any specific AMOEBA re-parametrization or ANI-2X dataset modification, the strategy required to either use of a tighter, but computationally inefficient Verlet/0.2ps integration or to perform an ANI-2X/AMOEBA BAR reweighting of a non-reactive AMOEBA set of trajectories, as discussed at the end of section 4.4. Due to the computational constraints, we chose the reweighting strategy that benefits from the efficiency of Tinker-HP to generate AMOEBA trajectories. Table S6-1 (SI) displays the ANI-2X/AMOEBA results obtained for C2 and C4. They are found in very good agreement with experiment with errors of 0.34 and 0.07 kcal/mol respectively. Again, the hybrid potential notably outperforms AMOEBA in these cases as ANI-2X clearly helps to improve accuracy for these two compounds. For the last ligand, C3, the nature of the problem appeared to be very different as the AMOEBA free energy prediction was almost perfect compared to experiment. In fact, we do not have a parametrization issue here and C3 represents the only case where a reactivity event occurred within our simulations. Indeed, when binding to the host, the C3 ligand adopts a cyclic conformation where its terminal OH and NH3 groups strongly interact. This is well captured by AMOEBA. Due to its reactive nature, the ANI-2X DNN potential is able to produce MD trajectories that include proton transfers between the groups suggesting that, for ANI-2X, the compound is actually a mix of two electronic states. As discussed in section 4.4, this situation is simply incompatible with an hybrid RESPA strategy. Again, we performed an ANI-2X/AMOEBA BAR reweighting computation using the well-defined initial AMOEBA electronic state to produce non-reactive classical trajectories. This led to a result apparently less in line with experiment than the AMOEBA one (1.76 kcal/mol vs 0.01 kcal/mol for AMOEBA) which was anticipated as ANI-2X tends to defavor the initial state. A solution would be to compute all possible states explored by ANI-2X/AMOEBA but the present discrepancy with experiment may be only apparent and the situation more complicated. Indeed, many things remain to be solved in the modeling of the SAMPL4 dataset. For example, in the SAMPL4 challenge overview, Muddana et al \cite{sampl_AMOEBA} reviewed the experimental conditions and concluded that it could be important to take into account the salt conditions and to go beyond the simple box neutralization. Indeed, in the event of a proton transfer, a new ionic specie being created, it would be interesting to study its interaction with different solutions of increasing ionic strength, especially in our case where the full simulation includes polarization effects. We have not done it at this stage as it would require a large number of additional simulations and we decided to keep the present C3 free energy prediction that could probably be improved in a forthcoming work. In any case, with C3, ANI-2X brings additional interpretative insights on the nature of the ligand. In a near future, it will be also interesting to investigate further the reactivity capabilities of ANI-2X/AMOEBA approach. Finally, it is worth noting that C3 is the weakest binder of the serie. ANI-2X/AMOEBA still predicts it as such in term of relative free energy of binding compared to the other compounds. 
  \\ 
Overall the hybrid ANI-2X/AMOEBA model results are in good agreement with experimental results, reaching, as for the solvation free energy studies, the chemical accuracy (average error of 0.94 kcal/mol vs experiment on the dataset) and  dividing by 2 the initial AMOEBA error. ANI-2X/AMOEBA can accurately predict binding free energies of flexible charged systems and the simulations clearly benefit from the addition of ANI-2X.

\section{Conclusion and Perspectives}
We first introduced Deep-HP, a novel massively parallel multi-GPU neural network platform which is a new component of the \TinkerHP\ molecular dynamics package. Deep-HP allows users to import their favorite Pytorch/TensorFlow Deep Neural Networks models within Tinker-HP. While Deep-HP enables to simulate million of atoms thanks to its MPI/Domain Decomposition setup, it introduces the possibility of reaching ns routine production simulation for hundreds of thousands of atoms biosystems with advanced neural network models such as ANI-2X. The platform capabilities have been demonstrated by simulating large biologically-relevant systems on up to \num{124} \GPU  with ANI-2X.\\
Since the platform allows the coupling of state-of-the-art polarizable force fields with any ML potential, we developed a new hybrid Deep Neural Networks/Polarizable potential that uses the ANI-2X ML potential for the solute-solute interactions and the AMOEBA polarizable force field for the rest. The development of the hybrid potential was motivated by the capability of AMOEBA to model accurately water-solute and water-water interactions whereas, a neural network such ANI is better able to capture complex intramolecular interactions at an accuracy approaching the CCSDT(T) gold standard of computational chemistry.\cite{ANI2X}  \\
We extended our hybrid model computational capabilities by designing RESPA-like multi-timestep integrators that can speedup simulations up to more than an order of magnitude with respect to Velocity Verlet \num{0.2}{fs}. In that context, the relative speedup of AMOEBA compared the hybrid ANI-2X/AMOEBA dropped from 40 to 2. The hybrid approach offers the inclusion of physically-motivated long-range effects (electrostatics and many-body polarization) and the capability to perform efficient Particle Mesh Ewald periodic boundary conditions simulations including polarizable counter ions. It also allows to benefit from the capability of the ANI-2X neural network to accurately describe the ligand potential energy surface leading to high-resolution exploration of its conformational space through the hybrid model MD simulation. The combination of these approaches allow to treat any type of ligands, including charged ones and opens the door to routine long timescale simulations using NNPs/PFFs up to million-atom biological systems, offering considerable speedup compared to traditional ligand binding QM/MM simulations.  \\
Our hybrid model accuracy was first assessed on solvation free energies of \num{71} molecules, with a large panel of different functional groups including charged ones, within three non-aqueous solvents and water.  The hybrid model outperforms AMOEBA accuracy on the non-aqueous solvents while performing almost equally-well in water, opening a path towards the simulation of complex biological processes for which the environment polarizability is important.\citep{D1SC00145K,interfacial_dina,D1SC05892D} 
We then reported the performance of our hybrid model on binding free energies of 14 host-guest challenging systems taken from the SAMPL host-guest binding competitions. Although most of ligands are charged, our hybrid model is able to outperform AMOEBA despite the complex chemical environments. ANI-2X/AMOEBA reached the chemical accuracy (average errors < 1kcal/mol w.r.t. experiment) on the testsets for both solvation and absolute binding free energies.\\ ANI-2X also provides new features such as the possibility to detect chemical modifications of the ligand thanks to the neural network reactive nature. In a near future, it would be interesting to systematically better converge the level of parametrization of AMOEBA and ANI-2X ligands in order to benefit from maximal multi-timestep acceleration. This should be easily achievable thanks to the recent improvements of the Poltype2 AMOEBA automatic parametrization framework.\cite{poltype2}  In this line, adaptive-timestep alternatives to multi-timestepping using Velocity Jumps \cite{jumps} would also be beneficial and are under investigation. These reactivity events also led us to introduce an accurate reweighting strategy. Since it is computationally efficient and avoid the costly computation of DNN gradients, it may become one of the strategy for free energy predictions. Further work will analyse the multiple possibilities of neural network reweighting setups in order to assess their computational efficiency.  \\
Overall, the Deep-HP platform, which takes advantage of state-of-the-art \TinkerHP\ \gpu\ code, was able to produce within a few days more than 10 $\mu$s of hybrid NNPS/PFFs molecular dynamics simulations which is, to our knowledge, the longest MD biomolecular study encompassing neural networks performed to date. Such performances should continue to improve thanks to further Deep-HP optimizations, TorchANI updates and \GPU\ hardware evolutions. 
Deep-HP will enable the implementation of the next generation of improved MLPs \cite{doi:10.1063/5.0083669,rezajooei2022neural} and has been designed to be a place for their further development. It will include direct neural networks coupling with physics-driven contributions going beyond multipolar electrostatics and polarization through inclusion of many-body dispersion models.\citep{MBD_pier, DNN_MBD} As Deep-HP's purpose is to push a trained ML/hybrid model towards large scale production simulations, we expect extensions of the present simulation capabilities to other class of systems towards materials and catalysis applications. Overall, Deep-HP allows the present ANI-2X/AMOEBA hybrid model to go a step further towards one of the grails of Computation Chemistry which is the unification within a reactive molecular dynamics many-body interaction potential of the short-range quantum mechanical accuracy and of long-range classical effects, at force field computational cost.

\section*{Author contributions statement}
T. J.I, O. A and T. P performed simulations; \\
O. A, O. I, T. J.I, L. L and T. P contributed new code;\\
L. L, O. I, T. J.I, P. R., T. P, J--P. P contributed new methodology; \\
T. J.I, L. L., P. R., J--P. P  contributed analytical tool; \\ 
T. J.I, L. L, O. I, P. R., H. G, J--P. P analyzed data.\\ 
T. J.I, L. L, T. P, H. G, O. I and J--P. P wrote the paper;\\
J--P. P designed the research.

\section*{Code availability}
Deep-HP is part of the \TinkerHP\ package which is freely accessible to Academics via GitHub : https://github.com/TinkerTools/tinker-hp

\section*{Conflicts of interest}
There are no conflicts to declare.

\section*{Acknowledgements}

This work has received funding from the European Research Council (ERC) under the European Union's Horizon 2020 research and innovation program (grant agreement No 810367), project EMC2 (JPP). Simulations have been performed at GENCI on the Jean Zay machine (IDRIS, Orsay, France) on grant no A0070707671 and at TGCC (Bruyères le Châtel, France) on the Irène Joliot Curie machine. The work performed by H.G. and O.I. (PI) was made possible by the Office of Naval Research (ONR) through support provided by the Energetic Materials Program (MURI grant no. N00014-21-1-2476). This research is part of the Frontera computing project at the Texas Advanced Computing Center. Frontera is made possible by the National Science Foundation award OAC-1818253.




\balance


\renewcommand{\bibname}{References}
\bibliographystyle{rsc} 
\bibliography{sample}

\end{document}






\newpage
\section{Study of the Deep-HP Platform Scalability}

\begin{figure}
\centering 
a) \hspace{4cm} b) 
\\
\subfloat{\label{a}\includegraphics[width=.5\linewidth]{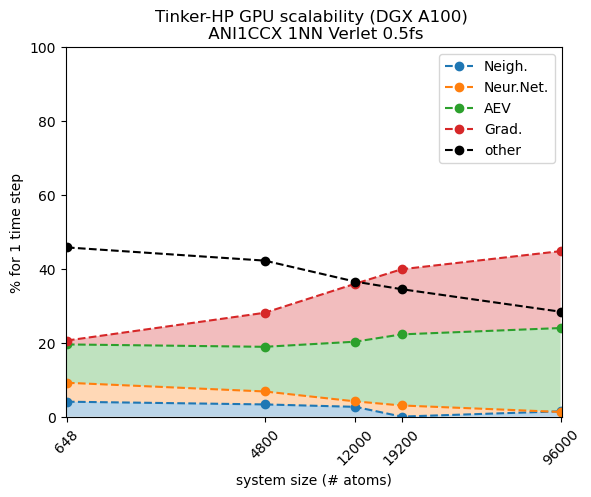}}\hfill
\subfloat{\label{b}\includegraphics[width=.5\linewidth]{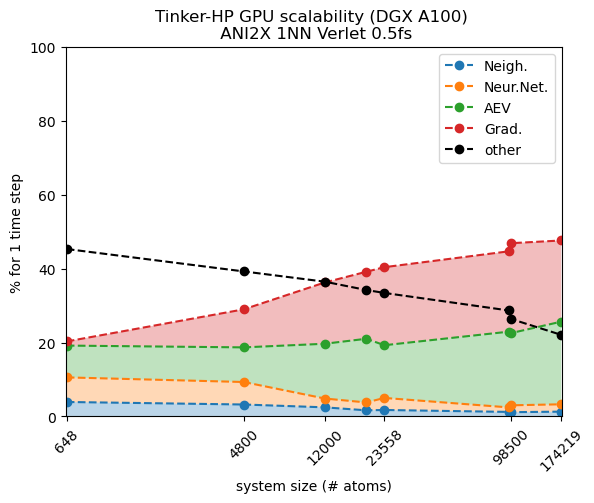}}\par
\subfloat{\label{c}\includegraphics[width=.5\linewidth]{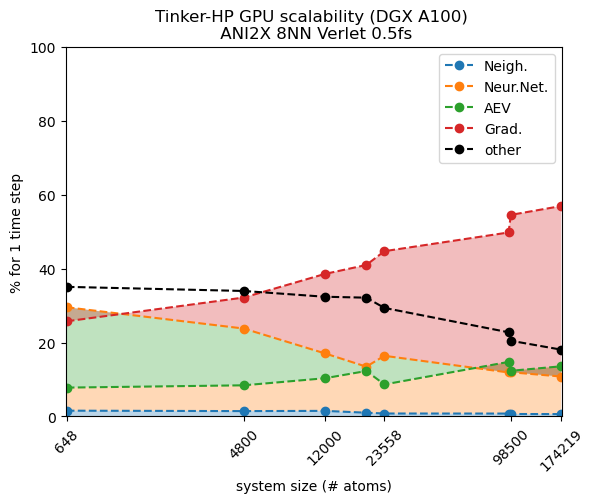}}
\caption{Performance plot comparison between ANI-2x(1NN), ANI-2x(8NN) and ANI-1ccx(1NN) for increasing system size on 1 GPU A100 in the NVE ensemble and using Verlet 0.5 fs time-step.}
\label{fig:rdf_ani}
\end{figure}
\newpage
\section{Hybrid Polarizable/ML potential model}
\subsection{Multi-timestep integrators: RESPA and RESPA1}

\begin{table*}[htp!]
\begin{center}
\noindent\makebox[\textwidth]{
\begin{tabular}{cccccccc} 
    \hline
    \rowcolor{Gray}
     & Exp. & AMOEBA & V(0.2) & R(0.25/1) & R(0.25/2) & R1*(0.25/2/4) & R1*(0.25/2/6) \\ \hline
    Benzene  & -0.87  & -0.37  & -0.83 & -0.97(-0.90) & \boldgreen{-0.87}(-0.88) & -1.69(-1.69) & -1.60 \\
    Water    & -6.32  & -7.35  & \boldgreen{-6.33}  & -6.29(-6.23) & -6.21(-6.22) & -6.39(\boldgreen{-6.33}) & X \\
\end{tabular}
}
\caption{\label{tab:delG_speedint}. Solvation free energy comparison for the benzene and water molecules. Comparison between experimental, AMOEBA and hybrid ANI-2X/AMOEBA results using Velocity Verlet, BAOAB-RESPA and BAOAB-RESPA1 integrators. The * corresponds to the use of hydrogen mass repartitioning (HMR). Simulations were performed in the NPT ensemble with 2ns and 5ns (in parentheses) BAR windows, with the BAOAB-RESPA/RESPA1 integrators.}
\end{center}
\end{table*}

\begin{table*}[htp!]
\begin{center}
\noindent\makebox[\textwidth]{
\begin{tabular}{ccccccccc} 
    \hline
    \rowcolor{Gray}
    {AMOEBA / ANI-AMOEBA} & ANI & {A/A(0.2)} & {A/A(0.25/1)} & {A/A(0.25/2)} & {A/A(0.25/2/4)} & {A/A(0.25/2/6)} \\ \hline
    AMOEBA V 0.2fs  & 10.78 & 8.62 & 1.81 & 1.02 & 0.61 & 0.48 \\
    AMOEBA V 1fs  & 44.64 & 35.69 & 7.52 & 4.24 & 2.53 & 1.96 \\
    AMOEBA R 0.25/2  & 49.50 & 39.58 & 8.35 & 4.70 & 2.80 & 2.18 \\
    Integrator-type  & V  & V  & R  & R  & R1 & R1(HMR)   \\ \bottomrule
\end{tabular}
}
\caption{\label{tab:speedint_AMOEBA}. Relative performances of usual AMOEBA setups: Velocity Verlet 0.2fs, Velocity Verlet 1fs and BAOAB-RESPA (0.25/2fs) compared to ANI alone (denoted ANI) and the hybrid ANI-2X/AMOEBA approach (denoted A/A) using several simulation setups for the solvated benzene system (3000 atoms).}
\end{center}
\end{table*}

\begin{table*}[htp!]
\begin{center}
\noindent\makebox[\textwidth]{
\begin{tabular}{ccc} 
    \hline
    \rowcolor{Gray}
    {A100} & {waterhuge (12000)} & {cox (174219)} \\ \hline
    ANI-2x V 0.2fs  & 1.02 & 0.084\\
    ANI-2x V 0.5fs  & 2.55 & 0.21 \\
    AMOEBA V 0.2fs  & 6.54 & 0.90 \\
    AMOEBA V 0.5fs  & 14.18 & 1.96  \\ \bottomrule
    Ratio V 0.2fs  & \textbf{6.41} & \textbf{10.71} \\
    Ratio V 0.5fs  & \textbf{5.56} & \textbf{9.33} \\ \bottomrule
\end{tabular}
}
\caption{\label{tab:speedvelo_AMOEBA}. Performance of AMOEBA and ANI-2x in ns per day using Velocity-Verlet (V) integrator with 0.2fs and 0.5fs time-step.}
\end{center}
\end{table*}
\newpage
\subsection{ANI-2X/AMOEBA Binding Free Energies through AMOEBA simulation}
In this section, we explain the basic principles of the reweighted Benett Acceptance Ratio (BAR) method. Let's denote by $i$ and $i+1$, two states for which we want to compute the free energy difference $\Delta F'_{i, i+1} = F'_{i+1}-F'_{i} = -\frac{1}{\beta}\ln \langle e^{-\beta (U'_{i+1}-U'_{i})}\rangle_{i}$ characterized by reduced energies $U_{i}$ and $U_{i+1}$. Common estimators such as BAR is based on the assumption of overlap meaning each configurations have non-negligible probability under the two states. For example, when performing alchemical free energies, the two states have different decoupling/scaling factor on the Van der Waals and electrostatic potentials. $\Delta F'_{i, i+1}$ can be estimate by using the energy differences, $\Delta U'_{i+1,i}$ sampled in state $i+1$, $\Delta U'_{i,i+1}$ sampled in state $i$, and the BAR estimator, the latter being constructed to minimize the variance and the mean-squared error. Within the targeted ensemble $U'$, the BAR estimator consists of solving self-consistently, over $k$: 
\begin{equation}
\begin{aligned}
    \Delta F'_{i, i+1}(k+1) &= -\frac{1}{\beta}\ln\frac{\langle 1/[1+\exp{\beta (U'_{i}-U'_{i+1}+C)]} \rangle_{i+1}'}{\langle 1/[1+\exp{\beta(U'_{i+1}-U'_{i}+C)]} \rangle_{i}'} + C \\
    & = -\frac{1}{\beta}\ln\frac{\frac{\int d^{N} \textbf{r} (1/[1+\exp{\beta(U'_{i}-U'_{i+1}+C)])} e^{-\beta (U'_{i+1})}}{\int d^{N} \textbf{r}  e^{-\beta (U'_{i+1})}}}{\frac{\int d^{N} \textbf{r} (1/[1+\exp{\beta(U'_{i+1}-U'_{i}+C)])} e^{-\beta (U'_{i})}}{\int d^{N} \textbf{r}  e^{-\beta (U'_{i})}}} + C
\end{aligned}
\end{equation}

with $C=\Delta F'_{i, i+1}(k)$ and $\beta$ the thermodynamic beta. \\
Now, let's suppose we have two ensemble $U$ and $U'$ with small overlap for each states $i$. If we want to obtain $\Delta F'_{i, i+1}$ from this other ensemble $U$, we can notice:

\begin{equation}
\begin{aligned}
    \langle 1/[1+\exp{\beta (U'_{i}-U'_{i+1}+C)]} \rangle_{i+1}' &=  \frac{\int d^{N} \textbf{r} (1/[1+\exp{(U'_{i}-U'_{i+1}+C)/RT])} e^{-\beta (U'_{i+1}-U_{i+1})} e^{-\beta (U_{i+1})}}{\int d^{N} \textbf{r}   e^{-\beta (U'_{i+1}-U_{i+1})} e^{-\beta (U_{i+1})}} \\
    & = \frac{\int d^{N} \textbf{r} (1/[1+\exp{(U'_{i}-U'_{i+1}+C)/RT])} e^{-\beta (U'_{i+1}-U_{i+1})} e^{-\beta (U_{i+1})}}{\int d^{N} \textbf{r} e^{-\beta (U_{i+1})}} \\
    &X\frac{\int d^{N} \textbf{r} e^{-\beta (U_{i+1})}}{\int d^{N} \textbf{r} e^{-\beta (U'_{i+1}-U_{i+1})} e^{-\beta (U_{i+1})}} \\
    & = \frac{\langle [\exp{\beta (U'_{i+1}-U_{i+1})}]/[1+\exp{\beta (U'_{i}-U'_{i+1}+C)]} \rangle_{i+1}}{\langle \exp{\beta (U'_{i+1}-U_{i+1})}\rangle_{i+1}}
\end{aligned}
\end{equation}

With Eq. (1), this gives the reweighted BAR formula, Eq. (3): 

\begin{equation}
\begin{aligned}
    \Delta F(k+1) = \frac{\langle [\exp{\beta (U'_{i+1}-U_{i+1})}]/[1+\exp{\beta (U'_{i}-U'_{i+1}+C)]} \rangle_{i+1}/\langle \exp{\beta (U'_{i+1}-U_{i+1})}\rangle_{i+1}}{\langle [\exp{\beta (U'_{i}-U_{i})}]/[1+\exp{\beta (U'_{i+1}-U'_{i}+C)]} \rangle_{i+1}/\langle \exp{\beta (U'_{i}-U_{i})}\rangle_{i}}
\end{aligned}
\end{equation}

Because it has been shown that AMOEBA and ANI-2X/AMOEBA share an overlap with numerous ligand, in the following we used $U'$ as a target hybrid ANI-2X/AMOEBA potential and $U$ as the AMOEBA reference. In addition to aleviate the intrinsic sampling problem of machine learning potential it allows to estimate the free energy of systems which may not be stable with the MLP during a dynamic, such as in our case the C2/C3/C4 host-guest complexes.  

\newpage
\subsection{Solvation free energies}
In the following, the experimental solvation free energies are taken from the Minnesota solvation database and from experimental solvation free energies taken from	Abraham	et al. The AMOEBA solvation free energies of Table S1-1, S1-2 were taken from Ren et al. (Poltype 2) and from Essex et al. for Table S2, S3, s4. \\ \\

\textbf{Table S1-1.} ANI-2X/AMOEBA calculated solvation free energies for small molecules in water ($\epsilon$=78.30) against AMOEBA and experimental data. The lowest absolute error is in green. ANI-2X/AMOEBA: RMSE: 0.70 kcal/mol; \boldgreen{21}/39. AMOEBA: RMSE: 0.59 kcal/mol; \boldgreen{18}/39. 
\begin{center}
\noindent\makebox[\textwidth]{
\begin{tabular}{cccccc} \toprule
    \multicolumn{4}{c}{{$\Delta G_{solv} (kcal/mol)$}} & \multicolumn{2}{c}{{$\Delta \Delta G_{solv} (kcal/mol)$}}\\ \midrule
    {Molecule} & {Exp} & {ANI/AMOEBA} & {AMOEBA} & {ANI/AMOEBA} & {AMOEBA} \\ \midrule
    Methylsulfide  & -1.24 & -0.87(-0.80)  & -0.83 & \boldgreen{0.37}(0.44) & 0.41 \\
    Methylethylsulfide  & -1.50  & -1.38(-1.33) & -1.28 & \boldgreen{0.12}(0.17) & 0.22  \\
    Water  & -6.32  & -7.28(-7.23) & -7.35  & 0.96(\boldgreen{0.91}) & 1.03 \\
    Trimethylamine  & -3.24  & -2.28(-2.16) & -2.36 & 0.96(1.08) & \boldgreen{0.88}   \\ 
    Toluene  & -0.89   & -0.35(-0.36) & -0.26  & 0.44(\boldgreen{0.43}) & 0.63 \\
    Propanol  & -4.83  & -5.20(-5.17) & -5.30 & 0.37(\boldgreen{0.34}) & 0.47  \\
    Propane  & 1.96  & 2.10(2.10) & 2.06 & 0.14(0.14) & \boldgreen{0.10}   \\
    Phenol  & -6.62  & -5.38(-5.45) & -5.48 & 1.24(1.17) & \boldgreen{1.14}   \\
    p-cresol  & -6.14  & -6.12(-6.13) & -6.27 & 0.02(\boldgreen{0.01}) & 0.13 \\
    n-butane & 2.08  & 2.20(2.01) & 2.23 & 0.12( \boldgreen{0.07}) & 0.15  \\
    Methylether & -1.90  & -1.13(-1.26) & -1.22 & 0.77(\boldgreen{0.64}) & 0.68 \\ 
    Methylamine & -4.56  & -2.96(-2.91) & -5.00 & 1.60(1.65) & \boldgreen{0.44} \\ 
    Methanol & -5.11  & -5.10(-5.08) & -5.14 & \boldgreen{0.01}(0.03) & 0.03 \\ 
    Methane & 1.99  & 1.68(1.68) & 1.73 & 0.31(0.31) & \boldgreen{0.27} \\ 
    Isopropanol & -4.76  & -6.11(-6.01) & -5.99 & 1.35(1.25) & \boldgreen{1.23} \\ 
    Imidazole & -9.63  & -10.12(-10.08) & -10.40 & 0.49(\boldgreen{0.45}) & 0.77 \\ 
    Hydrogensulfide & -0.44  & -0.06(-0.08) & -0.68 & 0.38(0.36) & \boldgreen{0.24} \\
    Ethylbenzene & -0.80  & -0.57(-0.48) & -0.16 & \boldgreen{0.23}(0.32) & 0.64 \\ 
    Ethylamine & -4.50  & -3.79(-3.93) & -4.08 & 0.71(0.57) & \boldgreen{0.42} \\ 
    Ethanol & -5.00  & -5.38(-5.19) & -5.39 & 0.38(\boldgreen{0.19}) & 0.39 \\ 
    Ethane & 1.83  & 1.98(1.93) & 1.82 & 0.15(0.10) & \boldgreen{0.01} \\ 
    Dimethylsulfide & -1.54  & -1.20(-1.26) & -1.25 & 0.34(\boldgreen{0.28}) & 0.29 \\
    Dimethylamine & -4.29  & -2.08(-2.07) & -3.86 & 2.21(2.20) & \boldgreen{0.43} \\
    Diethylsulfide & -1.60  & -1.10(-1.38) & -1.33 & 0.50(\boldgreen{0.12}) & 0.27 \\ 
    Benzene & -0.87  & -0.22(-0.52) & -0.37 & 0.65(\boldgreen{0.35}) & 0.50 \\ \bottomrule
\end{tabular}
}
\end{center}

\textbf{Table S1-2.} ANI-2X/AMOEBA calculated solvation free energies for small molecules in water ($\epsilon$=78.30) against AMOEBA and experimental data. The lowest absolute error are in green. ANI-2X/AMOEBA: RMSE: 0.70 kcal/mol; \boldgreen{13}/39. AMOEBA: RMSE: 0.59 kcal/mol; \boldgreen{25}/39.
\begin{center}
\noindent\makebox[\textwidth]{
\begin{tabular}{cccccc} \toprule
    \multicolumn{4}{c}{{$\Delta G_{solv} (kcal/mol)$}} & \multicolumn{2}{c}{{$\Delta \Delta G_{solv} (kcal/mol)$}}\\ \midrule
    {Molecule} & {Exp} & {ANI/AMOEBA} & {AMOEBA} & {ANI/AMOEBA} & {AMOEBA} \\ \midrule
    Aceticacid & -6.70  & -4.31(-4.24) & -4.95 & 2.39(2.46) & \boldgreen{1.75} \\ 
    Formicacid & -5.11  & -4.84(-4.96) & -4.98 & 0.27(0.15) & \boldgreen{0.13} \\ 
    Propylamine & -4.40  & -4.31(-4.42) & -4.63  & 0.09(\boldgreen{0.02}) & 0.23 \\ 
    Dimethyldisulfide & -1.83  & -0.49(-0.31) & -1.11 & 1.34(1.52) & \boldgreen{0.72} \\ 
    Di-n-butylamine & -3.24  & -2.18(-2.30) & -3.14 & 1.06(0.94) & \boldgreen{0.10} \\ 
    Methylisopropylether & -2.01  & -2.70(-2.52) & -2.15 & 0.69(0.51) & \boldgreen{0.14} \\ 
    Di-n-propylether & -1.16  & -1.65(-1.32) & -1.33 & 0.49(\boldgreen{0.16}) & 0.17 \\ 
    Methanethiol & -1.24  & -0.77(-0.83) & -0.83  & 0.47(\boldgreen{0.41}) & \boldgreen{0.41} \\ 
    Octan-1-ol & -4.09  & -5.36(-5.57) & -4.73  & 1.27(1.48) & \boldgreen{0.64} \\ 
    n-octane & 2.88  & 2.88(2.81) & 3.44 & \boldgreen{0.00}(0.07) & 0.56 \\ 
    Di-n-propylsulfide & -1.28  & -2.04(-1.71) & -0.53 & 0.76(\boldgreen{0.43}) & 0.75 \\ 
    Methylacetate & -3.13  & -2.41(-2.53) & -2.89 & 0.72(0.60) & \boldgreen{0.24} \\ 
    22-Dimethylbutane & 2.51  & 2.37(2.30) & 2.85 & \boldgreen{0.14}(0.21) & 0.34 \\ 
    n-butanethiol & -0.99  & -0.19(-0.34) & -0.48 & 0.80(0.65) & \boldgreen{0.51} \\ \bottomrule
\end{tabular}
}
\end{center}
\newpage
\textbf{Table S2.} ANI-2X/AMOEBA calculated solvation free energies for small	molecules in toluene ($\epsilon$=2.38) against AMOEBA and experimental data. The lowest absolute error are in green.  ANI-2X/AMOEBA: RMSE: 1.09/0.93(without methylamine) kcal/mol; \boldgreen{10}/20. AMOEBA: RMSE: 1.06/1.09(without methylamine) kcal/mol; \boldgreen{9}/20. 
\begin{center}
\noindent\makebox[\textwidth]{
\begin{tabular}{cccccc} \toprule
    \multicolumn{4}{c}{{$\Delta G_{solv} (kcal/mol)$}} & \multicolumn{2}{c}{{$\Delta \Delta G_{solv} (kcal/mol)$}}\\ \midrule
    {Molecule} & {Exp} & {ANI/AMOEBA} & {AMOEBA} & {ANI/AMOEBA} & {AMOEBA} \\ \midrule
    1,4-dioxane  & -4.91^{1} & -5.48 $\pm$ 0.02  & -5.43 $\pm$ 0.02 & 0.57 & \boldgreen{0.52} \\
    2-butanone  & -4.27  & -3.82 $\pm$ 0.02 & -3.74 $\pm$ 0.01  & \boldgreen{0.45} & 0.53  \\
    Acetic acid  & -4.00  & -3.28 $\pm$ 0.02 & -3.24 $\pm$ 0.04  & \boldgreen{0.72} & 0.76 \\
    Acetone  & -3.59  & -4.14 $\pm$ 0.02 & -4.09 $\pm$ 0.01  & 0.55 & \boldgreen{0.50}   \\ 
    Ammonia  & -2.38   & -0.45 $\pm$ 0.03 & 0.42 $\pm$ 0.03  & \boldgreen{1.93} & 2.80 \\
    Aniline  & -6.69  & -5.64 $\pm$ 0.02 & -5.11 $\pm$ 0.04 & \boldgreen{1.05} & 1.58  \\
    Ethanol  & -3.33  & -2.70 $\pm$ 0.02 & -2.60 $\pm$ 0.01  & \boldgreen{0.63} & 0.73   \\
    Methanol  & -2.18  & -2.05 $\pm$ 0.02 & -2.10 $\pm$ 0.04  & 0.13 & \boldgreen{0.08}   \\ 
    Methylamine  & -2.65  & -5.25 $\pm$ 0.05 & -2.88 $\pm$ 0.03  & 2.60 & \boldgreen{0.23} \\
    n-octane & -5.38  & -4.13 $\pm$ 0.04 & -3.82 $\pm$ 0.04  & \boldgreen{1.25} & 1.56  \\
    Nitromethane & -4.31  & -3.90 $\pm$ 0.02 & -4.00 $\pm$ 0.03  & 0.41 & \boldgreen{0.31} \\ 
    Phenol & -6.93  & -5.63 $\pm$ 0.02 & -5.33 $\pm$ 0.08  & \boldgreen{1.30} & 1.60 \\ 
    Pyridine & -5.13  & -4.74 $\pm$ 0.02 & -4.58 $\pm$ 0.04  & \boldgreen{0.39} & 0.55 \\ 
    Toluene & -5.12  & -3.97 $\pm$ 0.03 & -4.04 $\pm$ 0.03  & 1.15 & \boldgreen{1.08} \\ 
    Diethylamine & -3.75  & -3.23 $\pm$ 0.03 & -2.58 $\pm$ 0.04  & \boldgreen{0.52} & 1.17 \\ 
    Trimethylamine & -2.71  & -3.27 $\pm$ 0.03 & -3.33 $\pm$ 0.08  & \boldgreen{0.56} & 0.62 \\ 
    Hexanoic acid & -6.97  & -5.53 $\pm$ 0.03 & -5.83 $\pm$ 0004  & 1.44 & \boldgreen{1.14} \\ 
    Methylacetate & -3.81  & -3.74 $\pm$ 0.02 & -3.72 $\pm$ 0.06  & \boldgreen{0.07} & 0.09 \\ 
    Methylbenzoate & -7.96  & -6.53 $\pm$ 0.03 & -7.18 $\pm$ 0.03  & 1.43 & \boldgreen{0.78} \\ 
    Hydrogen peroxide & -3.14  & -2.31 $\pm$ 0.01 & -3.34 $\pm$ 0.05  & 0.83 & \boldgreen{0.20} \\ \bottomrule
\end{tabular}
}
\end{center}
\newpage
\textbf{Table S3.} ANI-2X/AMOEBA calculated solvation free energies, with BAR error, for small molecules in acetonitrile ($\epsilon$=36.64) against AMOEBA and experimental data. The lowest absolute error are in green. ANI-2X/AMOEBA: RMSE: 0.69 kcal/mol $\pm$ 0.02; \boldgreen{4}/6. AMOEBA: 0.71 kcal/mol $\pm$ 0.03; \boldgreen{2}/6. 
\begin{center}
\noindent\makebox[\textwidth]{
\begin{tabular}{cccccc} \toprule
    \multicolumn{4}{c}{{$\Delta G_{solv} (kcal/mol)$}} & \multicolumn{2}{c}{{$\Delta \Delta G_{solv} (kcal/mol)$}}\\ \midrule
    {Molecule} & {Exp} & {ANI/AMOEBA} & {AMOEBA} & {ANI/AMOEBA} & {AMOEBA} \\ \midrule
    1,4-dioxane  & -5.33^{1} & -5.64  & -5.55 & 0.31 & \boldgreen{0.22} \\
    2-butanone  & -4.73  & -3.74 & -3.78  & 0.99 & \boldgreen{0.95}  \\
    Ethanol  & -4.43  & -3.74 & -3.66  & \boldgreen{0.69} & 0.77   \\
    n-octane & -3.57  & -3.71 & -3.86  & \boldgreen{0.14} & 0.29  \\
    Nitromethane & -5.62  & -4.66 & -4.64  & \boldgreen{0.96} & 0.98 \\ 
    Toluene & -4.68  & -4.07 & -4.04  & \boldgreen{0.61} & 0.64 \\ \bottomrule
\end{tabular}
}
\end{center}
\newpage
\textbf{Table S4.} ANI-2X/AMOEBA calculated solvation free energies, with BAR error, for small	molecules in DMSO ($\epsilon$=47.24) against AMOEBA and experimental data. The lowest absolute error are in green. ANI-2X/AMOEBA: RMSE: 0.80 kcal/mol $\pm$ 0.03; \boldgreen{4}/6. AMOEBA: 1.21 kcal/mol $\pm$ 0.03; \boldgreen{2}/6. 
\begin{center}
\noindent\makebox[\textwidth]{
\begin{tabular}{cccccc} \toprule
    \multicolumn{4}{c}{{$\Delta G_{solv} (kcal/mol)$}} & \multicolumn{2}{c}{{$\Delta \Delta G_{solv} (kcal/mol)$}}\\ \midrule
    {Molecule} & {Exp} & {ANI/AMOEBA} & {AMOEBA} & {ANI/AMOEBA} & {AMOEBA} \\ \midrule
    1,4-dioxane  & -4.90^{1} & -5.40  & -5.27 & 0.50 & \boldgreen{0.37} \\
    2-butanone  & -4.23  & -3.05 & -2.87  & \boldgreen{0.96} & 1.36  \\
    Ethanol  & -5.25  & -4.59 & -4.48  & \boldgreen{0.66} & 0.77   \\
    n-octane & -2.84  & -2.80 & -1.04  & \boldgreen{0.04} & 1.80  \\
    Nitromethane & -5.66  & -4.45 & -4.56  & 1.21 & \boldgreen{1.10} \\ 
    Toluene & -4.42  & -3.85 & -3.09  & \boldgreen{1.03} & 1.33 \\ \bottomrule
\end{tabular}
}
\end{center}

\newpage
\textbf{Table S5.} ANI-2X/AMOEBA (A/A) absolute binding free energies for the G9 guest of the SAMPL6 challenge against AMOEBA and experimental data. Simulations performed in the NPT ensemble using an RESPA (0.2/2) integrator with 2ns BAR windows. \\$\Delta G_{bind} = \Delta G_{host-guest} - \Delta G_{solvation} + G_{correction}$ with $G_{correction}=1.7 kcal/mol$ \\
Experimental value: $\Delta G_{exp}$ = -8.68 kcal/mol. 
\begin{center} 
\noindent\makebox[\textwidth]{
\begin{tabular}{ccccc} \toprule
    {Model} & {$\Delta G_{bulk}$} & {$\Delta G_{host-guest}$} & {$\Delta G_{bind}$} & {$\Delta \Delta G$} \\ \midrule
    ANI/AMOEBA RESPA 2fs^{a} & 52.57  & -62.96 & -8.69 & \boldgreen{0.01} \\
    AMOEBA ref. & 3.27  & -13.79 & -8.82 & 0.14 \\\bottomrule
\end{tabular}
}
\end{center}

\newpage
\textbf{Table S6-1.} ANI-2X/AMOEBA absolute binding free energies for complexes of the SAMPL4 challenge against AMOEBA and experimental data. Simulations performed in the NPT ensemble using an RESPA (0.2/2) integrator with 2ns and 5ns (in parentheses) BAR windows. * (C10) uses RESPA (0.25/1). ** (C2, C3, C4) value obtained using an ANI-2X/AMOEBA reweighting of classical AMOEBA trajectories with 5ns BAR windows. For C3 we also performed 10 ns BAR windows (in parenthese). \\ $\Delta G_{bind} = \Delta G_{host-guest} - \Delta G_{solvation} + G_{correction}$ with $G_{correction}=6.245 kcal/mol$ \\
ANI-2X/AMOEBA: RMSE: 0.94 kcal/mol $\pm$ 0.05; \boldgreen{11}/13. AMOEBA: RMSE: 1.81 kcal/mol; \boldgreen{2}/13.
\begin{center}
\noindent\makebox[\textwidth]{
\begin{tabular}{ccccccc} \toprule
    \multicolumn{2}{c}{} & \multicolumn{3}{c}{{$\Delta G_{solv} (kcal/mol)$}} & \multicolumn{2}{c}{{$\Delta \Delta G_{sol   v} (kcal/mol)$}}\\ \midrule
    {Guest} & {Charge} & {Exp} & {ANI/AMOEBA} & {AMOEBA} & {ANI/AMOEBA} & {AMOEBA} \\ \midrule
    C1 & 2 & -9.90 & -9.35(-9.96)  & -12.27  & 0.55(\boldgreen{0.06}) & 2.37 \\
    C2** & 1 & -9.60 & -9.26 & -6.46  & \boldgreen{0.34} & 3.14   \\
    C3** & 1 & -6.60 & -4.84(-5.21) & -6.59  & 1.76 & \boldgreen{0.01}  \\
    C4** & 2 & -8.40  & -8.47 & -11.34  & \boldgreen{0.07} & 2.94 \\
    C6 & 1 & -7.90 & -7.33(-7.31) & -6.18  & \boldgreen{0.57}(0.59) & 1.72 \\
    C7 & 1 & -10.10 & -9.04(-9.75) & -10.49  & 1.06(\boldgreen{0.35}) & 0.39 \\
    C8 & 1 & -11.80 & -10.77(-11.10) & -11.84  & 1.03(0.70) & \boldgreen{0.04} \\
    C9 & 1 & -12.60 & -15.93(-15.17) & -15.42  & 3.33(\boldgreen{2.57}) & 2.82 \\
    C10* & 2 & -7.90 & -6.44(-6.44) & -5.06  & \boldgreen{1.46}(\boldgreen{1.46}) & 2.84 \\
    C11 & 1 & -11.10 & -11.25(-10.86) & -10.48  & \boldgreen{0.15}(0.24) & 0.62 \\
    C12 & 1 & -13.30 & -12.29(-13.20) & -12.11  & 1.01(\boldgreen{0.10}) & 1.19 \\
    C13 & 1 & -14.10 & -14.16(-13.63) & -13.92  & \boldgreen{0.06}(0.47) & 0.18 \\
    C14 & 1 & -11.60 & -13.21(-12.19) & -12.41 & 1.61(\boldgreen{0.59}) &  0.81 \\ \bottomrule
\end{tabular}
}
\end{center}

\newpage
\textbf{Table S6-2.} ANI-2X/AMOEBA absolute binding free energies for the C5 complex of the SAMPL4 challenge small molecules against AMOEBA and experimental data. Simulations performed in the NPT ensemble using a RESPA(0.2/2) integrator for ANI-2X/AMOEBA with 2ns or 5ns (in parenthesis) BAR windows.
\begin{center}
\noindent\makebox[\textwidth]{
\begin{tabular}{ccccccc} \toprule
    \multicolumn{2}{c}{} & \multicolumn{3}{c}{{$\Delta G_{solv} (kcal/mol)$}} & \multicolumn{2}{c}{{$\Delta \Delta G_{solv} (kcal/mol)$}}\\ \midrule
    {Guest} & {Charge} & {Exp} & {ANI/AMOEBA} & {AMOEBA} & {ANI/AMOEBA} & {AMOEBA} \\ \midrule
    C5 & 1 & -8.50 & -3.37(-2.11) & -3.37  & \boldgreen{5.13}(6.39) & \boldgreen{5.13} \\
    C5b & 2 & -8.50 & -3.69(-3.36) & -3.39  & \boldgreen{4.81}(5.14) & 5.11 \\
    C5b$\_$2 & 2 & -8.50 & -13.12(-14.66) & -  & 4.62(6.16) & - \\ \bottomrule
\end{tabular}
}
\end{center}